\documentclass[twocolumn]{aastex63}

\usepackage{textcomp}
\usepackage{graphicx}
\usepackage{natbib}

\shorttitle{Tilting Ice Giants with a Spin-Orbit Resonance}
\shortauthors{Rogoszinski and Hamilton}

\begin{document}

\title{Tilting Ice Giants with a Spin-Orbit Resonance}

\author{Zeeve Rogoszinski}
\affiliation{Astronomy Department  \\
University of Maryland \\
College Park, MD 20742, USA}

\author{Douglas P. Hamilton}
\affiliation{Astronomy Department  \\
University of Maryland \\
College Park, MD 20742, USA}

\email{zero@umd.edu, dphamil@umd.edu}


\begin{abstract}
	
	Giant collisions can account for Uranus's and Neptune's large obliquities, yet generating two planets with widely different tilts and strikingly similar spin rates is a low-probability event. Trapping into a secular spin-orbit resonance, a coupling between spin and orbit precession frequencies, is a promising alternative, as it can tilt the planet without altering its spin period. We show with numerical integrations that if Uranus harbored a massive circumplanetary disk at least three times the mass of its satellite system while it was accreting its gaseous atmosphere, then its spin precession rate would increase enough to resonate with its own orbit, potentially driving the planet's obliquity to 70\textdegree. We find that the presence of a massive disk moves the Laplace radius significantly outward from its classical value, resulting in more of the disk contributing to the planet's pole precession. Although we can generate tilts greater than 70\textdegree~only rarely and cannot drive tilts beyond 90\textdegree, a subsequent collision with an object about $0.5\,M_{\oplus}$ could tilt Uranus from 70\textdegree~to 98\textdegree. Minimizing the masses and number of giant impactors from two or more to just one increases the likelihood of producing Uranus's spin states by about an order of magnitude. Neptune, by contrast, needs a less massive disk to explain its 30\textdegree~tilt, eliminating the need for giant collisions altogether.\\
	
\end{abstract}

\section{Introduction} \label{sec:intro}

Gas accretion from the protoplanetary disk onto the forming giant planets supplies enough spin angular momentum to drive any primordial obliquities, the angle between the spin axis of the planet and the normal to its orbital plane, toward 0\textdegree. Instead, we observe a wide range of tilts, with Uranus's as the extreme case at 98\textdegree. The leading hypothesis for Uranus's large obliquity is multiple giant impacts \citep{1989Metic..24R.251B,1990Icar...84..528K,1992Icar...99..167S,1997P&SS...45..181P,2012Icar..219..737M,2015A&A...582A..99I,2018ApJ...861...52K,2019MNRAS.487.5029K}, which are expected during the early stages of planetary formation (e.g., formation of Earth's Moon; \cite{2001Natur.412..708C}); this model, however, has significant drawbacks, mainly that the impactors need to be near-Earth-sized. By contrast, Neptune's obliquity is only 30\textdegree, so a single impactor close to the mass of Mars could be responsible. 


If multiple giant collisions were responsible for the planets' obliquities, then we should observe additional signatures. For instance, we would expect Uranus's and Neptune's spin periods to differ significantly, but we instead observe only a 6\% difference ($T_{U}=17.2$ hr,  $T_{N}=16.1$ hr). These nearly identical spin periods imply a shared genesis, possibly similar to that of Jupiter and Saturn \citep{2018AJ....155..178B, 2018NatAs...2..138B}, with gas accretion as the dominant source of spin angular momentum. Furthermore, sudden changes to a planet's obliquity can disrupt or even destabilize its satellite system, and yet Uranus's regular satellites are very similar in relative sizes and spacings to the Galilean satellites. Neptune's satellites were disrupted by capturing Triton \citep{2006Natur.441..192A}, but it is likely that its primordial satellite system was somewhat similar to that of Uranus \citep{2017AJ....154..208R}. The resulting debris disk from a single giant impact to generate a tilt greater than 90\textdegree~would also tend to be orbiting retrograde \citep{2012Icar..219..737M}, and fine tuning is required to generate the prograde orbiting satellite system. Lastly, a giant impact would likely evaporate the ices from the ejecta debris disk \citep{2004A&A...413..373M}, suggesting rock-dominated compositions when in fact the satellites are abundant in water ice.

Extending the collisional model to Saturn introduces further complications, as the total mass of the impactors required to tilt Saturn to its current obliquity is between 6 and $7.2\,M_{\oplus}$ \citep{parisi2002model}. A promising alternative solution posits that Saturn is currently in a secular spin-orbit resonance with Neptune in which the precession frequencies of Saturn's spin axis and Neptune's orbital pole match. \cite{2004AJ....128.2510H} and \cite{2004AJ....128.2501W} show that the resonance can tilt Saturn from a primordial 0\textdegree~obliquity to its current 27\textdegree~as Neptune slowly migrates outward. This scenario preserves the spin period and satellite system of a planet as it slowly tips over \citep{1965AJ.....70....5G}, which would neatly sidesteps every issue with the giant impact model. But today, Uranus's and Neptune's spin precession frequencies are far slower than the fundamental orbital precession frequencies in our solar system precluding an active resonance \citep{1999ssd..book.....M,2006Icar..185..312B}. 

One way to facilitate a frequency match is to assume that Uranus originated between Jupiter and Saturn at around 7 au. With this assumption, the spin precession rate is fast enough to resonate with a planet like Neptune located beyond Saturn; however, as the precession timescales are long, the timescale required for Uranus to remain near 7 au exceeds several million years \citep{2018CeMDA.130...11Q,2018DDA....49P..10R,2019inprep}. \cite{2010ApJ...712L..44B} suggested that if Uranus once harbored a satellite larger than the Moon, it could augment the planet's gravitational quadrupole moment enough to speed up its spin precession frequency and generate a resonance on a timescale on the order of $10^{6}$ yr. However, this model suffers from the same problem as the giant impact hypothesis in that the moon would need to be implausibly large, placed at a large distance from the planet, and would also need to be removed without exciting the rest of the satellite system. 

An early circumplanetary accretion disk could also enhance the planetary system's bulge and speed up the planet's spin precession rate, at least for a few million years. The ice giants must have once had gaseous accretion disks, as 10\% of their mass is hydrogen and helium \citep{1995P&SS...43.1517P,2000P&SS...48..143P}. Additionally, circumplanetary disks are thought to be the birthplaces of the planet's regular satellites \citep{2002AJ....124.3404C, 2006Natur.441..834C, 2018ApJ...868L..13S}. As the circumplanetary disk survives for only a few Myr, a strong resonance would be required to tip Uranus. 

\section{Spin-Orbit Resonance}

An external torque from the Sun on an oblate planet causes slow uniform precession of the planet's spin axis about the normal to its orbital plane \citep{1966AJ.....71..891C}. Similarly, torques from the surrounding giant planets cause a planet's orbit to precess about the normal to the invariable plane. One of the simplest spin-orbit resonances occurs when these two precession frequencies are commensurate. Here the spin axis remains at a fixed angle relative to the pole of the planet's orbit, and the two vectors precess uniformly about the total angular momentum vector of the solar system. The normal to the total angular momentum vector is the invariable plane, and the angle between the projection of the two precessing vectors into the invariable plane is the resonance angle given by 
\begin{equation}\label{resarg}
\Psi = \phi_{\alpha} - \phi_{g}.
\end{equation}
Here $\phi_{\alpha}$ and $\phi_{g}$ are the longitudes measured from a reference direction to the projections of the spin axis and nodal pole onto the invariable plane, respectively \citep{2004AJ....128.2510H}. 

Capturing into a secular spin-orbit resonance requires not only nearly identical precession frequencies but also a configuration of coplanar vectors. Solutions for vector orientations that yield equilibria about which the resonance angle can librate are called ``Cassini states'' \citep{1966AJ.....71..891C,1969AJ.....74..483P,1975AJ.....80...64W,2004AJ....128.2501W}. This resonance argument (Equation \ref{resarg}) librates about Cassini state 2, which is an equilibrium point where the spin axis and orbital pole coprecess on opposite sides of the normal to the invariable plane. Following a resonance capture, the resonance angle varies slowly resulting in a gradual increase to the planet's obliquity if a dissipative force is active \citep{1974JGR....79.3375W}. The strongest such resonance occurs when the planet's spin precession frequency matches its own nodal precession frequency, which may be possible if the planet once harbored a massive circumplanetary disk. We therefore seek such cases for which it may be possible to tilt Uranus and Neptune within the lifetime of their circumplanetary disks. 

\cite{2018CeMDA.130...11Q} also included mean-motion terms in their resonance arguments, and they demonstrated that the corresponding torques from these terms can be as large as their secular counterparts. Planets located near a mean-motion resonance have altered orbital precession frequencies that might also induce a spin-orbit resonance \citep{2019NatAs...3..424M}. Since there is a rich variety of formation scenarios for Uranus and Neptune, with some starting in or eventually entering into different resonance chains with the other giant planets \citep[e.g.][]{2007AJ....134.1790M,2009A&A...507.1041M,2011AJ....142..152L,2012AJ....144..117N,2017AJ....153..153D}, it is not clear which, if any, of these resonances were important in the past. Accordingly, here we will only focus on the simpler commensurability between the two secular terms to demonstrate the ability of massive circumplanetary disks to enable large obliquity excitations.

\subsection{Spin-axis Precession}

The precession frequency of a planet's spin axis incorporates the torques from the Sun and any satellites on the central body \citep{1966AJ.....71..891C,1991Icar...89...85T}. If $\hat{\sigma}$ is a unit vector that points in the direction of the total angular momentum of the planetary system, then
\begin{equation}\label{diffeq}
\frac{d\hat{\sigma}}{dt} = {\alpha}(\hat{\sigma}\times\hat{n})(\hat{\sigma}\cdot\hat{n})
\end{equation}
where $\hat{n}$ is a unit vector pointing in the direction of Uranus's orbital angular momentum, and $t$ is time. Accordingly, the axial precession period is
\begin{equation}\label{period}
T_{\alpha}=\frac{2\pi}{{\alpha}\cos\epsilon},
\end{equation}
where $\cos\epsilon = \hat{\sigma}\cdot\hat{n}$.

If the satellites' orbits are prograde and nearly equatorial, and their masses are much less than that of the central body, then the spin axis precession frequency near 0\textdegree~is \citep{1975AJ.....80...64W,1991Icar...89...85T,2004AJ....128.2501W}
\begin{equation}\label{prec}
{\alpha} = \frac{3{n}^{2}}{2} \frac{{J}_{2} + q}{K\omega + l}.
\end{equation}
Here $n = (GM_{\odot}/r_{P}^{3})^{1/2}$ is the planet's orbital angular speed, $r_{P}$ is its distance to the Sun, $\omega$ is its spin angular speed, $J_{2}$ is its quadrupole gravitational moment, and $K$ is its moment of inertia coefficient divided by $M_{P}R_{P}^{2}$. The value of $K$ is relatively uncertain and is inferred from interior models. The parameter
\begin{equation}\label{quadrupole}
q\equiv\frac{1}{2} {\sum}_{i} ({M_{i}}/{M_{P}})({a_{i}}/{R_{P}})^{2}
\end{equation}
is the effective quadrupole coefficient of the satellite system, and
\begin{equation}
l\equiv{R}_{P}^{-2} {\sum}_{i} ({M_{i}}/{M_{P}})(GM_{P}a_{i})^{1/2}
\end{equation}
is the angular momentum of the satellite system divided by $M_{P}R_{P}^{2}$. The masses and semi-major axes of the satellites are $M_{i}$ and $a_{i}$. We can modify the satellite parameters to instead describe a disk by simply replacing the summation with an integral with $M_{i}$ interpreted as the mass of the ringlet with width $\Delta{a}$ at a distance $a_{i}$. The mass of each ringlet is therefore $M_{i}=2\pi a\,\Delta a\, \Sigma(a)$, with $\Sigma(a)$ as the surface density profile of the disk.

Equation \ref{prec} neglects the effects of other planets that would increase $\alpha$ by only $\sim M_{P}/M_{\odot}\sim$ 0.1\%. Only Uranus's major regular satellites contribute significantly to these quantities, so at the present day, we have $q=1.56\times10^{-2}$ and $l=2.41\times10^{-7}$ s$^{-1}$. Furthermore, $J_{2}=3.34343\times10^{-3}$ and $K\omega=2.28\times10^{-5}$ s$^{-1}$, so $K\omega>>l$ and $q=4.7J_{2}$. At its current $\epsilon=98$\textdegree~obliquity, Uranus's spin precession period is about $T_{\alpha}=210$ Myr (or $\alpha=0.0062$ arcsec yr$^{-1}$), and at near zero obliquity, $T_{\alpha}=29$ Myr (or $\alpha=0.045$ arcsec yr$^{-1}$). The mass of Uranus's current satellite system is about $1.05\times10^{-4}\,M_{U}$ or $9.1\times10^{21}$ kg, so a more massive circumplanetary disk would increase $q$ and $\alpha$ considerably, especially for a slowly spinning planet. 

\subsection{Orbital Pole Precession}\label{sec:orb_prec}

Torques from neighboring planets cause a planet's orbit to precess, and Uranus's current orbital precession period is 0.45 Myr or 64 times faster than its present-day $\alpha$. The orbital precession rate would be even faster in the presence of the massive circumstellar disk. If the density profile of the circumstellar disk is the minimum-mass solar nebula (MMSN), then the total mass of the disk would be about $M_{d}=10M_{J}$ \citep{1981PThPS..70...35H}. Raising the total orbiting mass of the solar system by an order of magnitude should also increase the orbital precession frequencies of all the planets by a similar amount \citep{1999ssd..book.....M}. 

A planet's orbital precession rate is determined by perturbations from sections of the disk both interior and exterior to the planet. Assuming the density of the circumstellar disk follows a power-law profile with index $\beta_{-}$ inside the planet's orbit and $\beta_{+}$ outside, the precession rate is negative and is given as $g=g_{-}+g_{+}+g_{p}$ with
\begin{equation}
g_{-} = -\frac{3}{4}n\left(\frac{2-\beta_{-}}{4-\beta_{-}}\right)\left(\frac{1-\eta_{-}^{4-\beta_{-}}}{1-\eta_{-}^{2-\beta_{-}}}\right)\left(\frac{M_{d,-}}{M_{\odot}}\right)\left(\frac{R_{o,-}}{r_{p}}\right)^{2},
\end{equation}
\begin{equation}
g_{+} = -\frac{3}{4}n\left(\frac{2-\beta_{+}}{-1-\beta_{+}}\right)\left(\frac{1-\eta_{+}^{-1-\beta_{+}}}{1-\eta_{+}^{2-\beta_{+}}}\right)\left(\frac{M_{d,+}}{M_{\odot}}\right)\left(\frac{r_{p}}{R_{o,+}}\right)^{3}
\end{equation}
where $g_{-}$ is the orbital precession rate induced from the interior disk, $g_{+}$ is from the exterior disk \citep[see Appendix \ref{nodalprec} for derivation]{2013ApJ...769...26C}, and $g_{p}$ is the contribution from the other giant planets \citep{1999ssd..book.....M}. Here $n$ is the mean motion of the planet, $M_{d,-}$ and $M_{d,+}$ are the masses of the circumstellar disk interior and exterior to the planet, $R_{o,-}$ and $R_{o,+}$ are the outer radii of each respective disk, and $\eta$ is the ratio of the inner and outer disk radii. 

To calculate $g$, we set $r_{p}$ to be 19 au, and the inner and outer radii of the solar system to be 0.1 and 100 au. The index $\beta=1.5$ for a MMSN if the planets were formed near their current locations, and $\beta\approx2.2$ if the planets abide by the Nice model \citep{2007ApJ...671..878D}. For this range of $\beta$, assuming $\beta=\beta_{+}=\beta_{-}$ Uranus's orbital precession rate is faster than its current rate by a factor of 3--7. Here the contributions from the other giant planets to Uranus's orbital precession rate are minor, as the mass of the circumstellar disk is much larger than the forming giant planet cores. However, since Uranus and Neptune are categorically gas-limited, the ice giants likely were actively accreting their atmospheres only when the circumstellar disk was significantly depleted \citep{2017AJ....154...98F}. At this point in time, Jupiter and Saturn had almost finished forming, so $g_{p}$ is close to Uranus's current rate. Adding a depleted $1M_{J}$ circumstellar disk to the mostly formed planetary system would therefore increase Uranus's current orbital precession frequency by about 30\% - 60\%.

Capturing into a spin-orbit resonance requires that the orbital precession rate $g\approx\alpha\cos\epsilon$. We increase Uranus's current orbital precession rate by 30\% and vary the planetary and disk parameters to find solutions for Uranus's spin precession rate that yield resonances. If Uranus's orbital precession rate was faster, then the planet would need a more massive circumplanetary disk to increase its spin precession rate and generate a resonance. As Uranus accretes matter, its spin angular momentum will also increase, so, all else being equal, $\alpha$ will tend to decrease (Equation \ref{prec}). We therefore seek cases where $\alpha\cos\epsilon$ was initially larger than $g$ so that the system will pass through the resonance. If the masses of both circumplanetary and circumstellar disks deplete at the same rate, then both precession rates ($g$ and $\alpha\cos\epsilon$) decrease at similar rates, and capturing into resonance is difficult \citep{2019ApJ...876..119M}. We instead expect the two frequencies to change at different rates, especially as the planet's spin precession rate will increase as it builds up its circumplanetary disk. A slow spin rate and a massive circumplanetary disk are optimal for speeding up a planet's spin precession rate, but is this enough to generate a strong and lasting spin-orbit resonance as the planet grows? In the next section, we explore the conditions necessary for the planetary system to develop a resonance and tilt over on a million-year timescale. 

\section{Planet and Disk Conditions for Resonances}

Resonances can occur at many stages during the formation of ice giants, but sustaining a resonance long enough to substantially tilt a planet requires certain conditions to be met. We propose that circumplanetary disks can satisfy those conditions, and here we will discuss how disks form and how the planet evolves while accreting from a disk. 

\subsection{Growing Ice Giants}

The classic gas giant formation process can be broken into three stages \citep{1986Icar...67..391B,1996Icar..124...62P,2009Icar..199..338L}. In stage 1, the core forms from the aggregation of pebbles and planetesimals. During stage 2, core accretion slows as the planetary core exceeds several Earth masses and becomes capable of capturing an atmosphere as its escape velocity exceeds the thermal velocity of the nearby gas. The planet distorts the surrounding disk as it accretes, and the corresponding gravitational torques lead to shocked wave fronts that carve out a gap \citep{1979MNRAS.186..799L, 1980ApJ...241..425G,2015ApJ...807L..11D}. The gas flows from the circumstellar disk onto a circumplanetary envelope or disk before accreting onto the planet. The transition between planar disks and more spherical envelopes depends on the planet's temperature: the hotter the planet, the greater the thermal pressure and the more spherical the circumplanetary gas is \citep{2016MNRAS.460.2853S}. For Uranus and Neptune near their current locations, this transitional planet temperature is about 500 K \citep{2018ApJ...868L..13S}. Since modeling the planet's equation of state during formation extends beyond the scope of this paper, we will instead use simple growth models and disk profiles to approximate the planet's evolution.

Stage 2 lasts a few Myr as the planet slowly accretes gas and planetesimals. Once the protoplanet's gaseous atmosphere becomes more massive than its core, the planet undergoes runaway gas accretion (stage 3), and it can gain about a Jupiter's worth of mass in just 10,000 yr. There are several competing explanations for why Uranus and Neptune have not accreted enough gas to achieve runaway gas accretion. The standard explanation by \cite{1996Icar..124...62P} suggests that Uranus and Neptune were not able to accrete enough solids near their current locations before the entire protoplanetary disk dissipated. Pebble accretion, however, reduces the ice giants' growth timescale and allows gas giants to form more rapidly at greater distances \citep{2012A&A...544A..32L}, but when the core is massive enough to gravitationally perturb the surrounding gas disk, it creates a pressure barrier to isolate it from further pebble accretion \citep{2014A&A...572A..35L}. Alternatively, \cite{1999Natur.402..635T,2002AJ....123.2862T,2003Icar..161..431T} posited that Uranus and Neptune were formed between Jupiter and Saturn, and that Jupiter's and Saturn's cores happened to be more massive, allowing them to accrete most of the surrounding gas. When the solid-to-gas ratio in the circumstellar disk reached unity, there was not enough gas to damp the eccentricities and inclinations of the growing protoplanets. As such, dynamical instability is then triggered, and the ice giants are scattered outward. 

These models all assume Uranus and Neptune formed in a massive circumstellar disk. \cite{2017AJ....154...98F} argued that if Uranus's and Neptune's cores were formed close to Jupiter and Saturn later in solar system evolution, then the ice giants could have accreted their atmospheres in an already depleted circumstellar disk after they had been scattered and reached close to their current locations. If only 1\% ($\sim0.1M_{J}$) of the original circumstellar disk remained after the cores migrated outward, then there would have been just enough gas near the ice giants to form their atmospheres. This reduction implies a gas accretion duration for Uranus and Neptune on the order of $10^{5}$ yr given a nominal gas loss rate of $7\times10^{-10}M_{\odot}\,$yr$^{-1}$ \citep{2005MNRAS.358..283A}, but 2D and 3D gas accretion models suggest that some gas also crosses through the gap, bypassing the planets altogether \citep{1999ApJ...514..344B,2012ApJ...747...47T,2018AJ....155..178B}. Therefore, if less than half of the gas within the planet's vicinity is actually accreted \citep{2014Icar..232..266M,2018A&A...619A.165C}, then there needed to have been more gas to compensate, and we could expect a longer gas accretion timescale perhaps closer to 1 Myr. 

Regardless of how they formed, Uranus and Neptune would have had to harbor a circumplanetary disk at some point. This disk will at least initially maintain a steady state, but as the circumstellar disk dissipates, we expect the circumplanetary disk to disappear as well. We will therefore explore these two basic scenarios.

\subsection{Spin Evolution of Giant Planets} \label{sec:spin}

Circumplanetary disks regulate not only the growth rate of giant planets but also their spin rates  \citep{1986Icar...67..391B, 2009Icar..199..338L, 2010AJ....140.1168W}. We expect the planets to be spinning at near-breakup velocities if we only consider the hydrodynamics arising in an inviscid thin disk. We instead observe the giant planets, including the first giant exoplanet with a measured spin rate, $\beta$ Pictoris b \citep{2014Natur.509...63S}, spinning several times slower than their breakup rates. Thus, there must be some mechanism responsible for removing excess angular momentum. The solution may be a combination of magnetic braking caused by the coupling of a magnetized planet to an ionized disk \citep{2011AJ....141...51L, 2018AJ....155..178B}, polar inflows and additional outflows from a thick disk profile \citep{2012ApJ...747...47T}, and magnetically driven outflows \citep{1998ApJ...508..707Q,2003A&A...411..623F}; regardless, gas accretion is a significant source of angular momentum. It is therefore possible that the planets' spin rates prior to gas accretion were indeed slow, especially if their cores were made up of the accumulation of many small bodies striking randomly at the planet's surface \citep{1991Icar...94..126L, 1993Icar..103...67D, 1993Sci...259..350D, 1999Icar..142..219A}, but pebble accretion may also contribute a significant amount of prograde spin \citep{2020Icar..33513380V}. 

Since Uranus and Neptune spin at about the same rate and have similar gas content, we suspect that gas accretion is the primary source of their respective spin periods. We model the effect of gas accretion on the planet's spin state by incrementally adding angular momentum to the planet according to
\begin{equation}\label{eq:angmom}
\vec{l}_{gas}=\Delta M R_{P}V_{orbit}\lambda\,\hat{z},
\end{equation}
where $\Delta M$ is the differential mass of the gas accreted at that time step, $V_{orbit}=\sqrt{GM_{P}/R_{P}}$ is the circular velocity at the edge of the planet, $M_{P}$ and $R_{P}$ are the mass and radius of the planet, and $\hat{z}$ points normal to the orbital plane. The planet's spin rate then grows as $L/(KMR^{2})$. Since accretion is not 100\% effective, we include the constant $\lambda$ with $\lambda<1$. The accretion efficiency is relatively unconstrained, and, in practice, we tune $\lambda$ so that Uranus's final spin angular momentum matches its current value. We will therefore explore a range of initial spin rates, from where the planet is spinning fast enough such that its spin angular momentum is close to its current value to cases where the planet is initially spinning slower than that.

Finally, we assume that the angular momentum transport to the planet is smooth, even when the circumplanetary disk is warped at high planetary obliquities. Planets accrete gas from a circumplanetary disk driven by accretion mechanisms such as magnetorotational instability-triggered turbulent viscosity \citep{1973A&A....24..337S,1991ApJ...376..214B} or shock-driven accretion via global density waves \citep{2016ApJ...832..193Z}. Tilting the planet with a quadrupole torque presents unique challenges to the accretion mechanism, as additional wavelength disturbances are introduced when the disks are warped \citep{1995ApJ...438..841P}. Since we fix the accretion rate to 1 $M_{\oplus}$ per million years, the details of the accretion mechanism are relatively unimportant. Furthermore, as the dominant accretion mechanism in these systems is unknown, we use our fiducial constant surface density profile. \cite{2014MNRAS.441.1408T} showed a big dip in the disk's density near the Laplace radius if the viscosity is low, but the disk remains unbroken if the viscosity increases. This dip is more pronounced at higher obliquities, yet the authors show that warped disks remain intact even at $\epsilon=60$\textdegree-- $\!$70\textdegree. Circumplanetary disks can also tear if the density is too low, but the resulting instabilities and momentary variations to the accretion rate occur over short timescales \citep{2018MNRAS.476.1519D}. Global disk properties, such as the average accretion rate, remain mostly unaffected.

\subsection{Laplace Radius} \label{sec:laplace}
The outer edges of circumplanetary disks are not well known, but estimates place them somewhere between 0.3 and 0.5 Hill radii  \citep{1998ApJ...508..707Q,2009MNRAS.397..657A,2012MNRAS.427.2597A,2009MNRAS.392..514M,2010AJ....140.1168W,2011MNRAS.413.1447M,2012ApJ...747...47T,2014ApJ...782...65S,2016ApJ...832..193Z}; however, only a portion of the disk will tilt with the planet. This region is located within the planet's Laplace radius, or warping radius, which is the transition point where perturbations from the planet are comparable to those from the Sun. Orbits well beyond a planet's Laplace radius precess about the ecliptic, while orbits well inside this point precess about the planet's equator. The Laplace radius, which also discriminates regular from irregular satellites, is approximately
\begin{equation}\label{laplace}
R_{L} \approx \left(2J_{2,tot}\frac{M_{P}}{M_{\odot}}R_{P}^{2}r_{P}^{3}\right)^{1/5}
\end{equation}
\citep{1966RvGSP...4..411G,2008ssbn.book..411N,2016Natur.539..402C}. For reference, Uranus's current Laplace radius is about 76.5 Uranian radii, and without the effect of the satellite system, it reduces to 54 $R_{U}$. 


Here, $J_{2,tot}$ is the total quadrupole moment of the planetary system, or the sum of the quadrupole moment of the planet ($J_{2}$) and the disk ($q$). The planet's $J_{2}$ depends quadratically on the planet's spin rate,
\begin{equation}\label{j2}
J_{2} \approx \frac{\omega^{2} R_{P}^{3} k_{2}}{3GM_{P}}
\end{equation}
\citep{2009ApJ...698.1778R}, where $k_{2}$ is the Love number. The Love number is a dimensionless parameter that characterizes a planet's susceptibility to tidal deformation, and the larger the number, the greater the bulge. A more slowly spinning planet has a larger $\alpha$, but also a smaller $J_{2}$ and hence a smaller Laplace radius, which may limit the disk's contribution to the planet's quadrupole moment. Furthermore, the planet may have had an initially smaller $K$, the planet's dimensionless moment of inertia, as the planet was hot and puffy. This means also having a smaller Love number (see Figure 4.9 of \citep{1999ssd..book.....M}) but also a larger spin rate for a given mass, radius and angular momentum.

The disk mass contained within the Laplace radius determines the disk's gravitational quadrupole moment $q$. If the surface density profile of the disk falls as a power law and $q\gg J_{2}$, then we can transform Equation \ref{laplace} to be approximately
\begin{equation}\label{laplace2}
R_{L} \approx \left(\frac{2\pi\Sigma_{0}R_{o}^{\beta}r_{P}^{3}}{(4-\beta)M_{\odot}}\right)^{1/(1+\beta)}
\end{equation}
where $\Sigma_{0}$ is the central surface density of the disk, $R_{o}$ is the outer radius of the disk, and $\beta>0$ is the power-law index (see Appendix \ref{laplacederiv} for derivation). The Laplace plane transition from the planet's equator to the ecliptic is actually a continuous curve, but a sharp transition at $R_{L}$ where everything inside it tilts in unison is a sufficient approximation. The disk's contribution to $q$ has a stronger dependence on $a$ than the disk's mass, and we find that $q$ can be dozens of times larger than $J_{2}$ for a range of disk sizes. Therefore, we can easily excite the planet's spin precession frequency to values much greater than the planet's nodal precession rate, and as the disk dissipates, we can achieve a spin-orbit resonance. 


\section{Changing the Obliquity of a Growing Protoplanet}

A massive circumplanetary disk is capable of increasing a planet's spin precession rate and generating a resonance, and in this section, we will investigate how massive this disk needs to be. We will first explore how the spin precession frequency changes for different disk profiles and then expand our model by having the planet also evolve with the disk.

\subsection{Constant Surface Density Profile} \label{sec:const}

After the planet opens up a gap, gas flows from the circumstellar disk and concentrates near the planet's centrifugal barrier. This is the gas's pericenter distance, where the centrifugal force is balanced by the planet's gravitational pull. The gas then heats up and spreads, forming a compact Keplerian rotating disk \citep{2009MNRAS.392..514M}. Calculations for the average specific angular momentum of the gas are calibrated for Jupiter and Saturn, but when adopting Lissauer's (\citeyear{1995Icar..114..217L}) analytic estimate of the disk's specific angular momentum to Uranus, the disk extends to about $60\,R_{U}$. This fiducial radius for Neptune is $100\,R_{N}$ because the planet is located farther away from the Sun. To simplify, we will assume a constant surface density profile, which is possible for a low planetary accretion rate $\dot{M}$ \citep{2016ApJ...832..193Z}. A portion of the disk extends beyond the centrifugal barrier, puffing up to smoothly connect with the circumstellar disk. The surface density in this outer region falls off with increasing distance as a power law. If the planet is larger than its centrifugal barrier, then this is the only part of the disk.

\begin{figure}[t]
	
	\centering
	\includegraphics[width=0.5\textwidth]{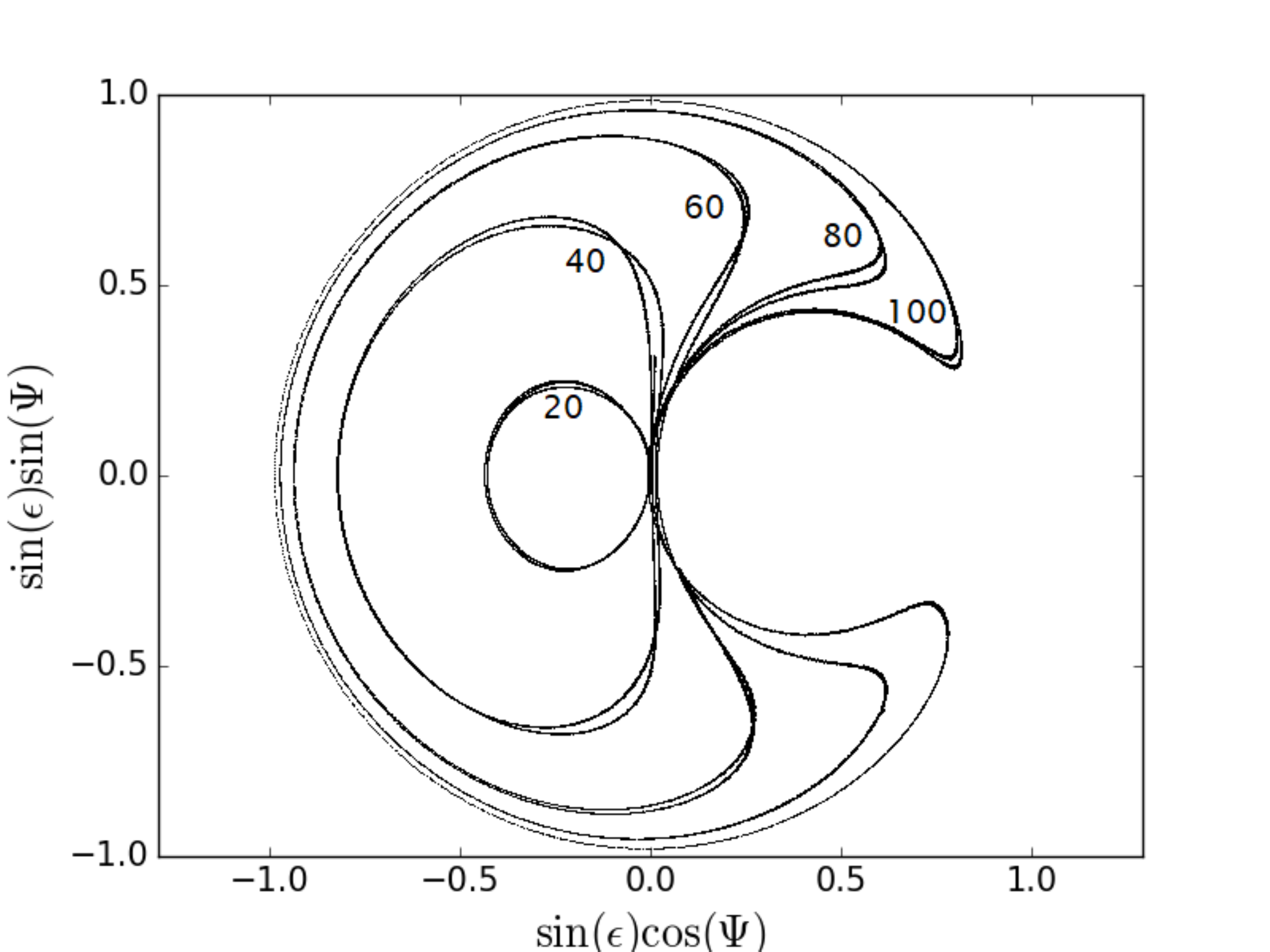}
	
	\caption{Evolution of the resonance angle $\Psi$ and obliquity $\epsilon$ for static disks with different disk masses. The resonance angle librates about the equilibrium point indefinitely when trapped into resonance; otherwise, the resonance angle circulates through a full $2\pi$ radians. Each contour corresponds to a resonance trapping for different disk masses displayed in units of $M_{s}$, where $M_{s}=10^{-4}M_{U}$. Uranus's orbital precession rate in a solar system that includes a depleted circumstellar disk is about $2\times10^{-5}$ yr$^{-1}$, and the planet's spin precession rates near 0\textdegree~with 20 $M_{s}$ and 100 $M_{s}$ circumplanetary disks are $\alpha=1.4\times10^{-5}$ yr$^{-1}$ and $3.8\times10^{-5}$ yr$^{-1}$, respectively. If the mass of the disk increases well beyond $100\,M_{s}$, the planet's spin precession frequency will be too fast to allow a resonance capture. }
	\label{fig:polar}
\end{figure}

We track the motions of the planets using HNBody \citep{2002DDA....33.0802R} and then evaluate Equation \ref{diffeq} using a fifth-order Runge-Kutta algorithm \citep{1992nrca.book.....P, 2019inprep}. If we place Uranus at its current location and set its physical parameters to its current values, then a disk of constant density extending to $54\,R_{P}$, which is also the Laplace radius without the disk's influence, needs more than 20 times the mass of its satellite system (where $M_{s}=10^{-4}M_{U}$) to generate a spin-orbit resonance (Figure \ref{fig:polar}). Here the amplitude of the resonance angle increases for increasing disk masses, and, similar to first-order mean-motion resonances, the resonance center shifts locations as the distance from the resonance changes \citep{1999ssd..book.....M}. Larger disks will require less mass to generate a resonance, but they could extend beyond the classical Laplace radius, which we will discuss later. If Uranus's orbital precession rate is faster by a factor of 2 due to torques from a remnant solar nebula, then we will need twice as much mass to generate a spin-orbit resonance (Equation \ref{prec}). For comparison, \cite{2018ApJ...868L..13S} favored slightly smaller satellite disk masses of $M_{d}\approx10^{-3}M_{U}$. 

\subsection{A Shrinking Disk}

\begin{figure}[h]
	\centering
	\begin{tabular}[b]{@{}p{0.45\textwidth}@{}}
		\centering\includegraphics[width=0.5\textwidth]{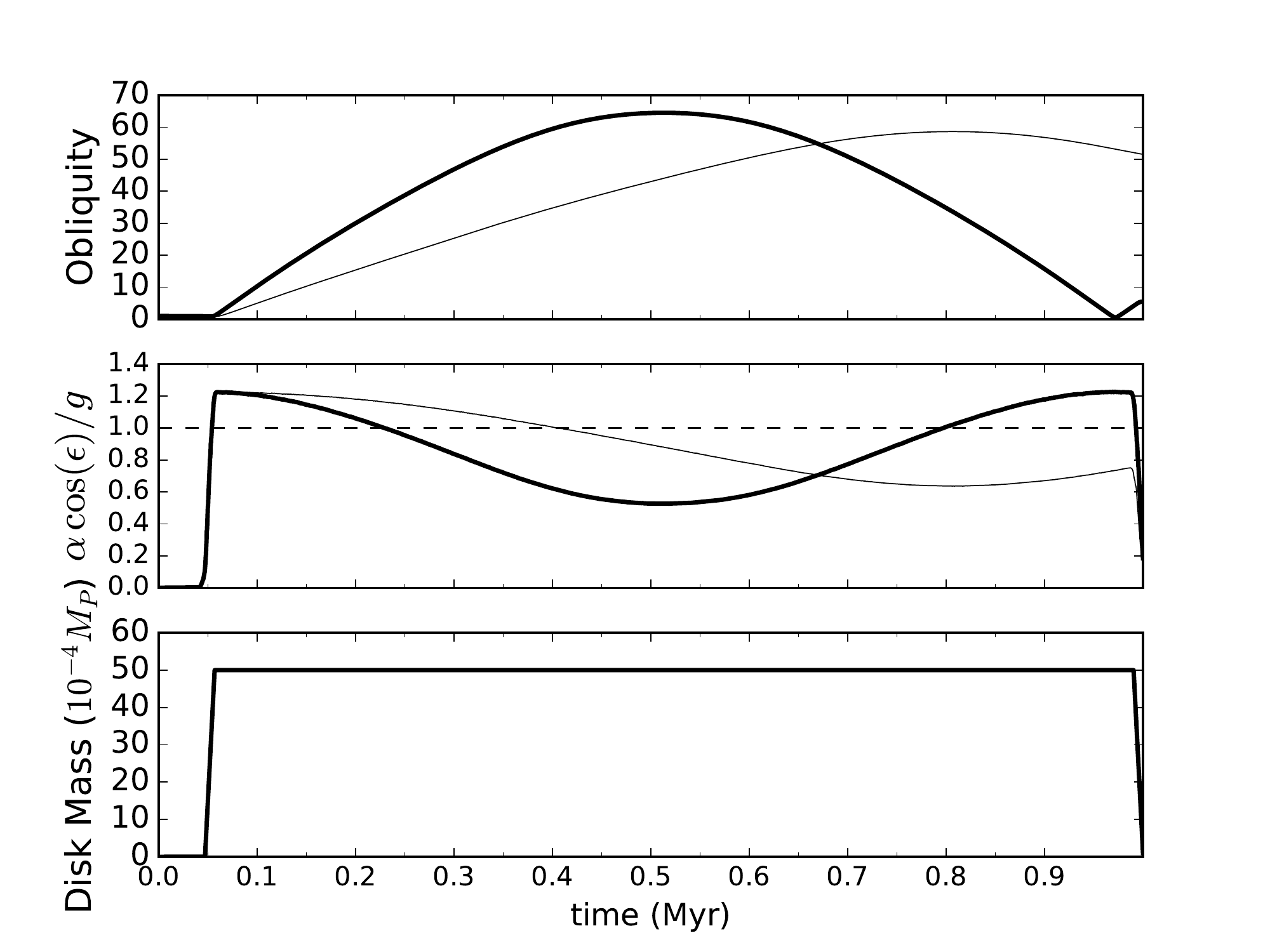} \\
		\centering (a)
	\end{tabular} \\
	\begin{tabular}[b]{@{}p{0.45\textwidth}@{}}
		\centering\includegraphics[width=0.5\textwidth]{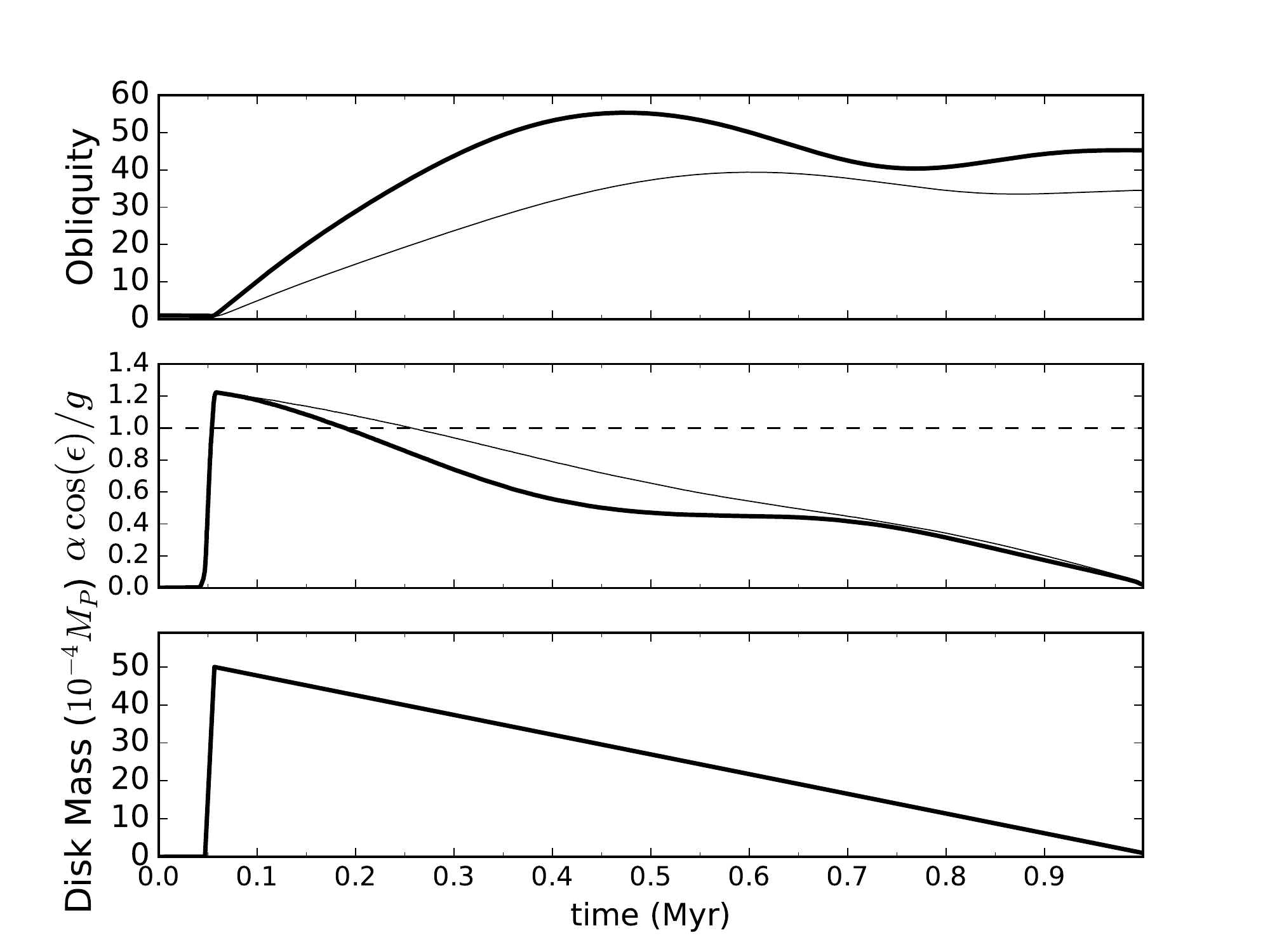} \\
		\centering (b)
	\end{tabular}

	\caption{(a) Uranus at its current state but surrounded by a $50\,M_{s}$ constant density disk for a duration of about 1 Myr. The disk extends all the way to $54\,R_{U}$. Thick black lines assume that Uranus's inclination is $i_{U}=10$\textdegree~while thin lines indicate $i_{U}=5$\textdegree. The top panel shows the evolution of the planet's obliquity in degrees; the middle panel shows the evolution of the precession frequencies, with the dashed line indicating the resonance location; and the bottom panel shows the evolution of the mass of the disk. (b) Same scenario, but the disk's mass decreases over time.}
	\label{fig:disc1}
\end{figure}

The circumplanetary disk will evolve as the planet accretes, and the spin precession rate will vary depending on how the disk changes. The ice giants need to accrete about $1\,M_{\oplus}$ of gas in 1 Myr, so at a constant accretion rate of $1\,M_{\oplus}$ Myr$^{-1}$ the lifetime of the gas is $\tau_{d}= M_{d}/\dot{M} \sim 10^{4}$ yr, or a tiny fraction of the accretion time span. We can therefore expect a sharp initial rise to the mass of the disk, and then either the disk maintains that mass in a steady state \citep{2016ApJ...832..193Z,2018ApJ...868L..13S} or it steadily decreases as the circumstellar disk dissipates. Figure \ref{fig:disc1} shows the evolution of the resonance for both cases. In this set of figures, Uranus's physical parameters are tuned to their current values, and we place a $50\,M_{s}$ disk around the planet to augment the planet's spin precession rate to generate a spin-orbit resonance. Here we see that a circumplanetary disk in steady state is capable of driving obliquities about 15\% higher than disks that deplete over time. This is because the resonance frequency decreases as the disk shrinks, which limits the amount of time the planet can be nearly resonant. Finally, a larger orbital inclination will drive obliquities to higher degrees on shorter timescales as the resonance is stronger. 

\subsection{Setting the Orbital Inclination}

The strength of the resonance is proportional to the planet's orbital inclination \citep{2004AJ....128.2510H}, so it takes longer to drive Uranus to higher obliquities in a resonance capture for low $i_{U}$. The evolution of the planets' orbital inclinations are unknown, but planet-planet interactions \citep{2008ApJ...678..498N} or mean-motion resonances \citep{2003ApJ...597..566T} can amplify a planet's inclination, which can then damp through dynamical friction as the planet migrates outward. Scattering small particles, such as circumstellar gas or planetesimals, places them on high-velocity orbits, and in response, the planet's orbit circularizes and flattens. For simplicity, we require the planet to maintain a constant orbital inclination for the entire duration of the simulation. This is justified because the damping timescale in a depleted gaseous disk is greater than 1 Myr, and it is even longer for planetesimal scattering. 

\begin{figure}[h]
	
	\centering
	\includegraphics[width=0.5\textwidth]{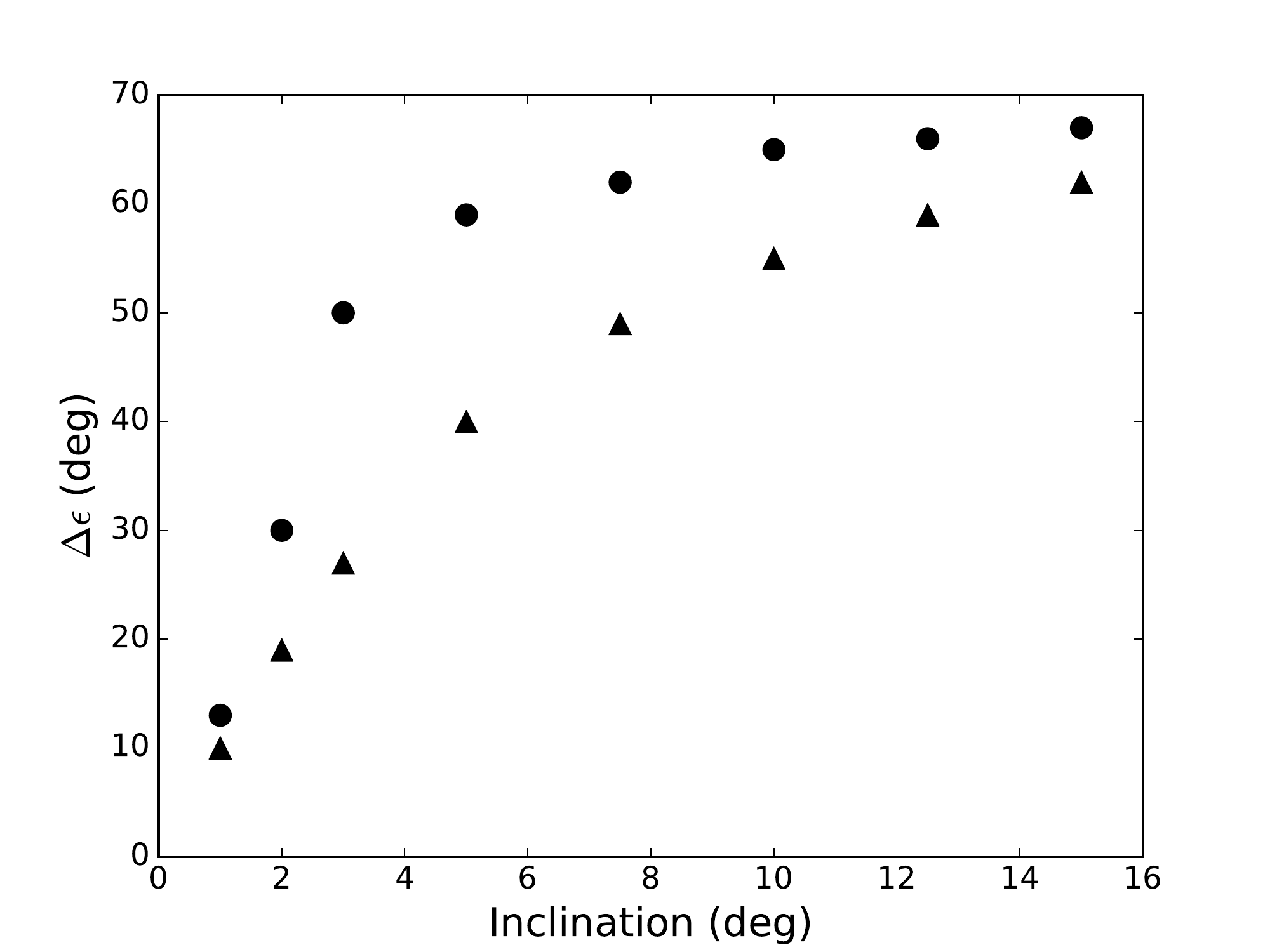}
	
	\caption{Maximum degree of tilting for a range of orbital inclinations if the disk's mass remains constant (circles) or is decreasing (triangles). The planet and disk possesses the same physical characteristics as described in Figure \ref{fig:disc1}, and the duration of each simulation is 1 Myr. For reference, Uranus's current inclination relative to the solar system's invariable plane is about 1\textdegree. }
	\label{fig:inc}
\end{figure}

Figure \ref{fig:inc} summarizes the maximum change in Uranus's obliquity for a suite of numerical simulations like that displayed in Figure \ref{fig:disc1} with different assumed inclinations. If the disk maintains a constant mass, then the planet can undergo a resonance capture for inclinations above about 5\textdegree. Extending the duration of the simulation in Figure \ref{fig:inc} from 1 to 10 Myr can yield resonance captures for orbits with inclinations closer to 2\textdegree. While resonance captures are capable of driving obliquities to higher values, the planet's final obliquity could be less than maximum. This is because while the resonance is active, the planet's obliquity oscillates as the spin axis librates. The resonance for a depleting disk, on the other hand, will last only briefly as a resonance kick, and in this case, the planet's final obliquity will remain fixed after the resonance terminates. Regardless, we can achieve substantial tilts if the planet's orbital inclination was greater than 5\textdegree. 

\subsection{Growing Uranus and tilting it over} 

\begin{figure}
	\centering
	\begin{tabular}[b]{@{}p{0.45\textwidth}@{}}
		\includegraphics[width=0.5\textwidth]{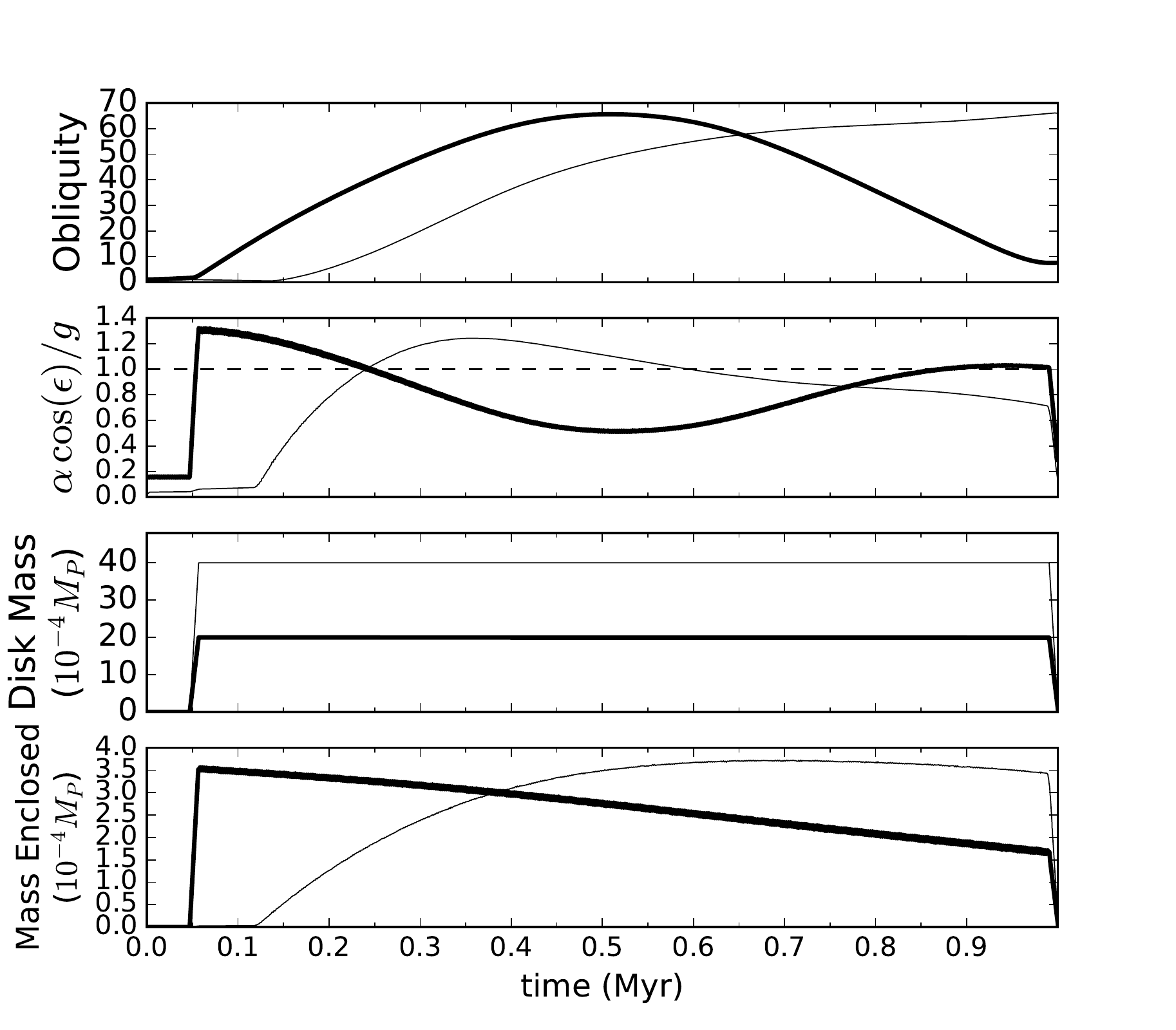} \\
		\centering (a)
	\end{tabular} \\
	\begin{tabular}[b]{@{}p{0.45\textwidth}@{}}
		\includegraphics[width=0.5\textwidth]{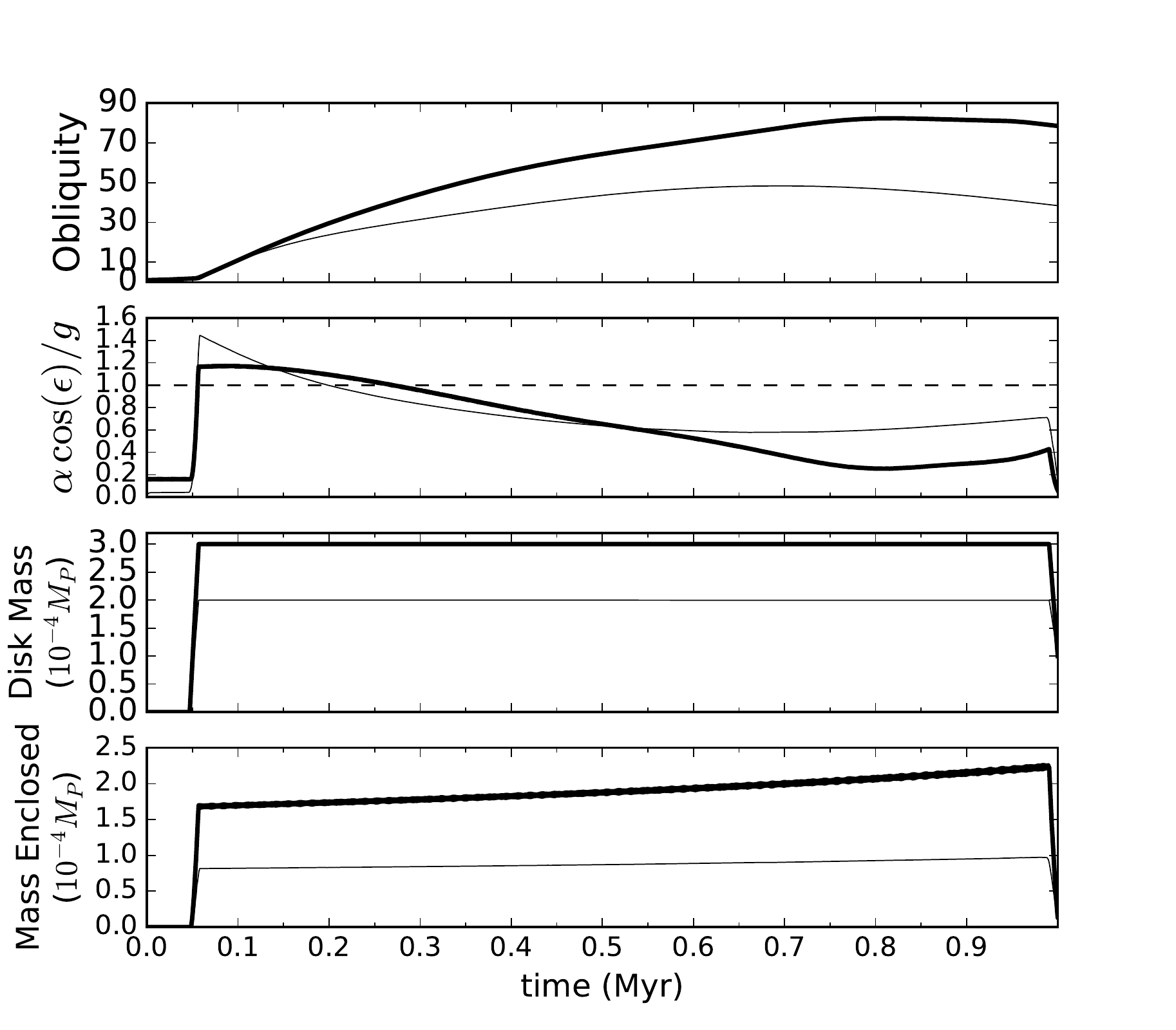} \\
		\centering (b)
	\end{tabular}
	
	\caption{(a) Evolution of Uranus's obliquity for a growing planet where the Laplace radius is determined by only by the evolution of the planet's $J_{2}$. The planet's mass grows from $0.9$ to $1.0\,M_{U}$, and the radius grows from $80$ to $120\,R_{U}$. The circumplanetary disk extends to 0.5 Hill radii, and the surface density falls by 3 orders of magnitude. The thick bold lines have a Uranus initial angular momentum $L_{0}$ of approximately the planet's current value, $L_{U}=1.3\times10^{36}$ kg m$^{2}$ s$^{-1}$, and the thin bold lines have $L_{0}\approx 0.25\,L_{U}$. In the former case, $R_{L}$ ranges from 130 to 140 $R_{U}$, while for the latter, it ranges from 80 to 140 $R_{U}$. The results for having $L_{0}\approx 0.25\,L_{U}$ do not noticeably change if the planet's initial spin angular momentum is lower. In both cases, Uranus's orbital inclination is set to 10\textdegree. The bottom panel shows the disk mass contained within Uranus's Laplace radius, which contributes to the pole precession rate $\alpha$. (b) Same situation, but $R_{L}$ grows according to Equation \ref{laplace2}. Here the circumplanetary disk extends to 0.1 Hill radii, consistent with \cite{2018ApJ...868L..13S}, and $R_{L}\approx200\,R_{U}$.}
	\label{fig:tilt2}
\end{figure}

In the last section, we investigated how to generate a resonance by changing disk properties. Here we explore how the planet's spin precession rate and obliquity evolve as Uranus accretes its atmosphere and grows. After core accretion stops, Uranus acquires a 1$M_{\oplus}$ atmosphere over roughly 1 million yr. Its radius is initially large ($\sim80\,R_{U}$), as the planet is hot from the energy added to it from accreting planetesimals \citep{1986Icar...67..391B,1996Icar..124...62P,2009Icar..199..338L}. The radius grows exponentially and terminates at around $120\,R_{U}$, when the gas fully dissipates. The angular momentum of the planet also grows as the planet accretes gas, so the planet's spin rate varies as $L/(KMR^{2})$, with the caveat that the planet's final angular momentum does not exceed its current value (Equation \ref{eq:angmom}). Finally, a disk with an extended density profile will mostly contribute to the planetary system's quadrupole moment, and $R_{L}$ increases according to Equation \ref{laplace2}. The other physical limit is a thin disk in which $R_{L}$ depends only on the planet's $J_{2}$, which results in a much smaller Laplace radius. We will display both cases in the following runs. 

With a growing planet, even a constant disk mass lasting over 1 Myr can generate a resonance capture (Figure \ref{fig:tilt2}), and, for a planet with an initial spin angular momentum close to its current value, the disk needs to be have $M_{d}=3\times10^{-4}-2\times10^{-3}\,M_{U}$ to tilt the planet. Recall that \cite{2018ApJ...868L..13S} calculated a circumplanetary disk around Uranus of about $10^{-3}\,M_{U}$ which falls comfortably within this mass range. In the case where $R_{L}$ changes according to Equation \ref{laplace2} (Figure \ref{fig:tilt2}b), a less massive disk is needed if the planet's spin rate was slower since $\alpha \propto q/K\omega$. Here we can tilt Uranus's obliquity all the way to 80\textdegree, though in most cases, it reaches about 50\textdegree.

If we instead artificially keep the Laplace radius small by having it depend only on the planet's $J_{2}$, as in Figure \ref{fig:tilt2}a, then the size of the Laplace radius eventually decreases relative to the size of the planet. Assuming angular momentum is conserved, the spin rate falls as $R_{P}^{2}$ as the planet grows, and using Equations \ref{laplace} and \ref{j2}, we find $R_{L}/R_{P} \propto R_{P}^{-4/5}$. As a result, for an initially fast-spinning planet, both the quadrupole moment of the disk and the planet's spin precession rate shrink. A more massive disk is needed if the planet was initially spinning slowly in order to compensate for a small Laplace radius earlier in the planet's evolution. In this case, the Laplace radius initially grows as the planet spins up, and, as represented by the thin bold line in the bottom panel of Figure \ref{fig:tilt2}a, more of the disk's mass is enclosed. At around 0.6 Myr, the size of the Laplace radius compared to the size of Uranus begins to shrink because the planet's spin angular momentum is nearing its current value. These figures show that the quadrupole moment of the disk cannot be neglected; its primary effect is to reduce the amount of mass needed in the disk by about an order of magnitude. We find that a disk mass of $4\times10^{-3}\,M_{U}$ is more than sufficient to generate a spin-orbit resonance.

\begin{figure}
	\centering
	\begin{tabular}[b]{@{}p{0.45\textwidth}@{}}
			\includegraphics[width=0.5\textwidth]{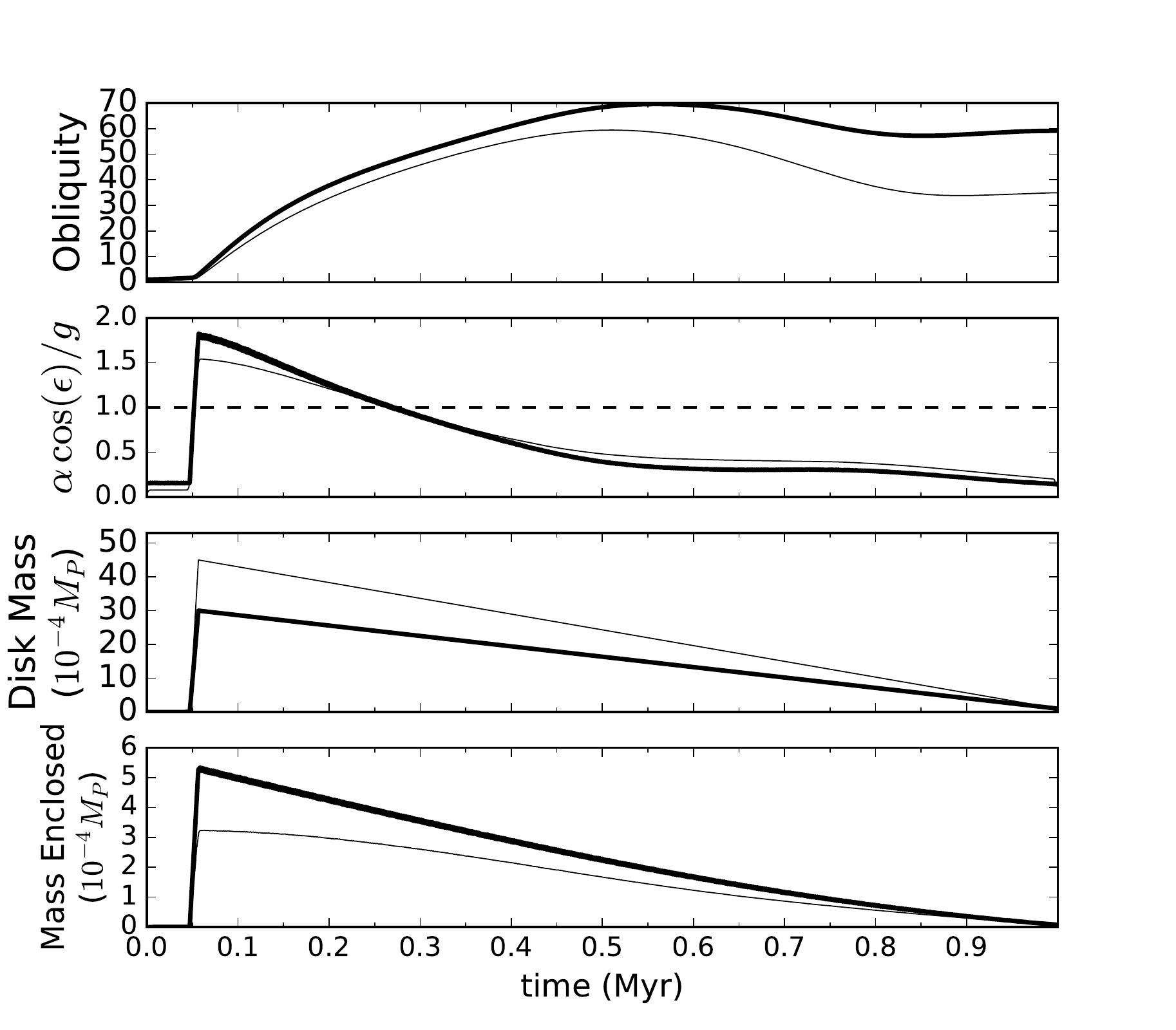} \\
		\centering (a)
	\end{tabular} \\
	\begin{tabular}[b]{@{}p{0.45\textwidth}@{}}
			\includegraphics[width=0.5\textwidth]{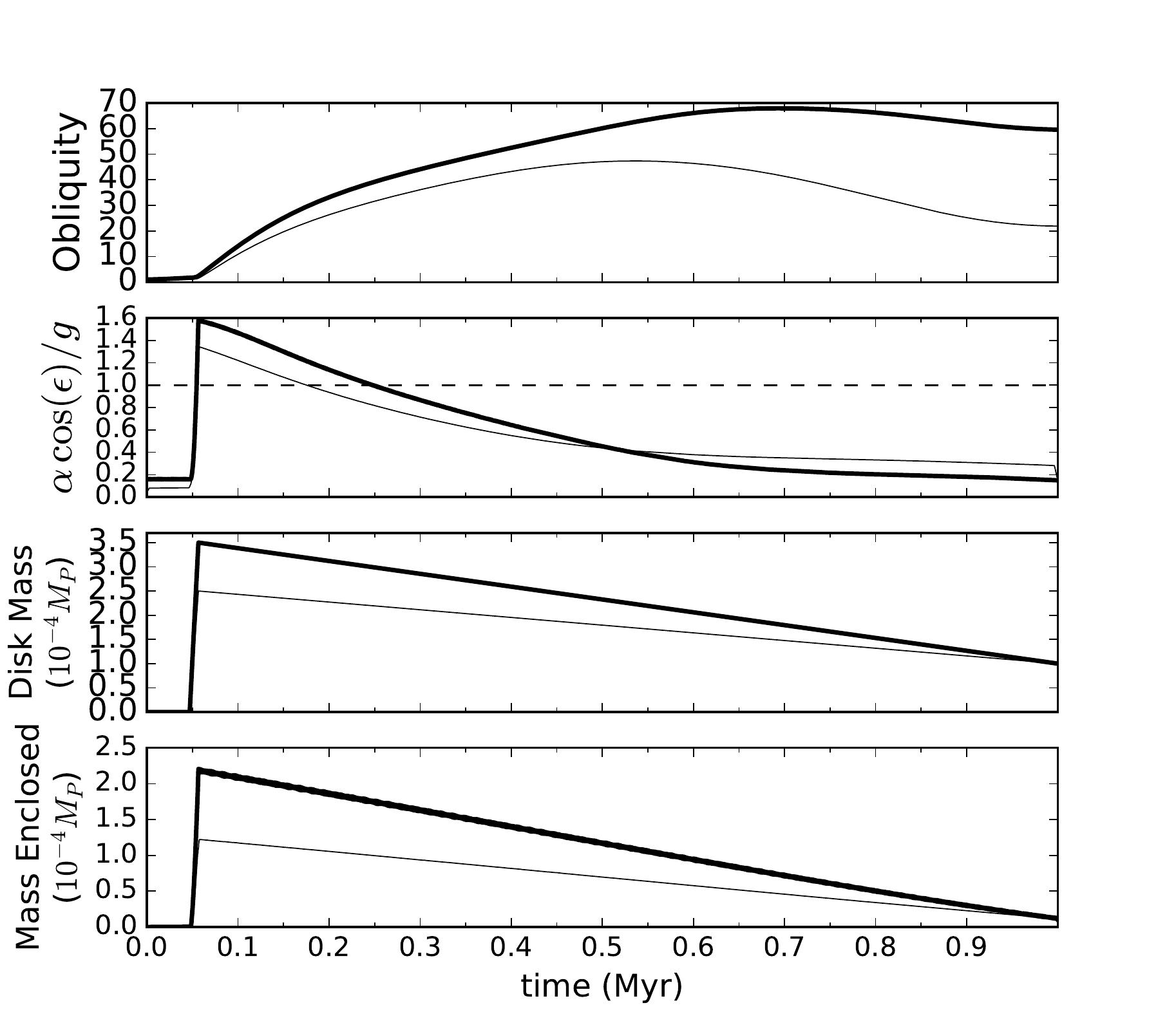} \\
		\centering (b)
	\end{tabular}
	\caption{(a) Same situation as in Figure \ref{fig:tilt2} but with the circumplanetary disk's mass decreasing over time. Here the thick bold lines have a Uranus initial angular momentum $L_{0}$ of approximately the planet's current value, while the thin bold lines have $L_{0}\approx 0.5\,L_{U}$. For the $L_{0}\approx L_{U}$ case, $R_{L}$ ranges from 130 to 145 $R_{U}$, while for $L_{0}\approx 0.5\,L_{U}$ it ranges from 80 to 140 $R_{U}$. (b) Same situation, but $R_{L}$ grows according to Equation \ref{laplace2}, and $R_{L}\approx200\,R_{U}$.}
	\label{fig:tilt1}
\end{figure}

Figure \ref{fig:tilt1} instead depicts a depleting circumplanetary disk with an initial mass $M_{d}=2.5\times10^{-4}-4\times10^{-3}\,M_{U}$, and the planet evolves similarly to those shown in Figure \ref{fig:tilt2}. Regardless of how large $R_{L}$ is, the planet's spin precession frequency will decrease as $M_{d}$ decreases, and we can tilt Uranus to as high as 70\textdegree~for similarly sized disks, as in the constant disk mass case. As in Figure \ref{fig:tilt2}, we see that the disk's effect on the Laplace radius reduces the disk mass required for resonance by about a factor of 10. How the disk evolves for an already depleted circumstellar disk is likely more complicated than these idealized scenarios, but in the realistic scenarios depicted in Figures \ref{fig:tilt2}(b) and \ref{fig:tilt1}(b) Uranus requires a disk a few times the mass of the satellite system to be contained within $R_{L}$ to generate spin-orbit resonance. As such, a resonance is very possible, even with a circumplanetary disk concentrated close to the planet.

\subsection{Tilting Neptune}

\begin{figure}
	\centering
	\includegraphics[width=0.5\textwidth]{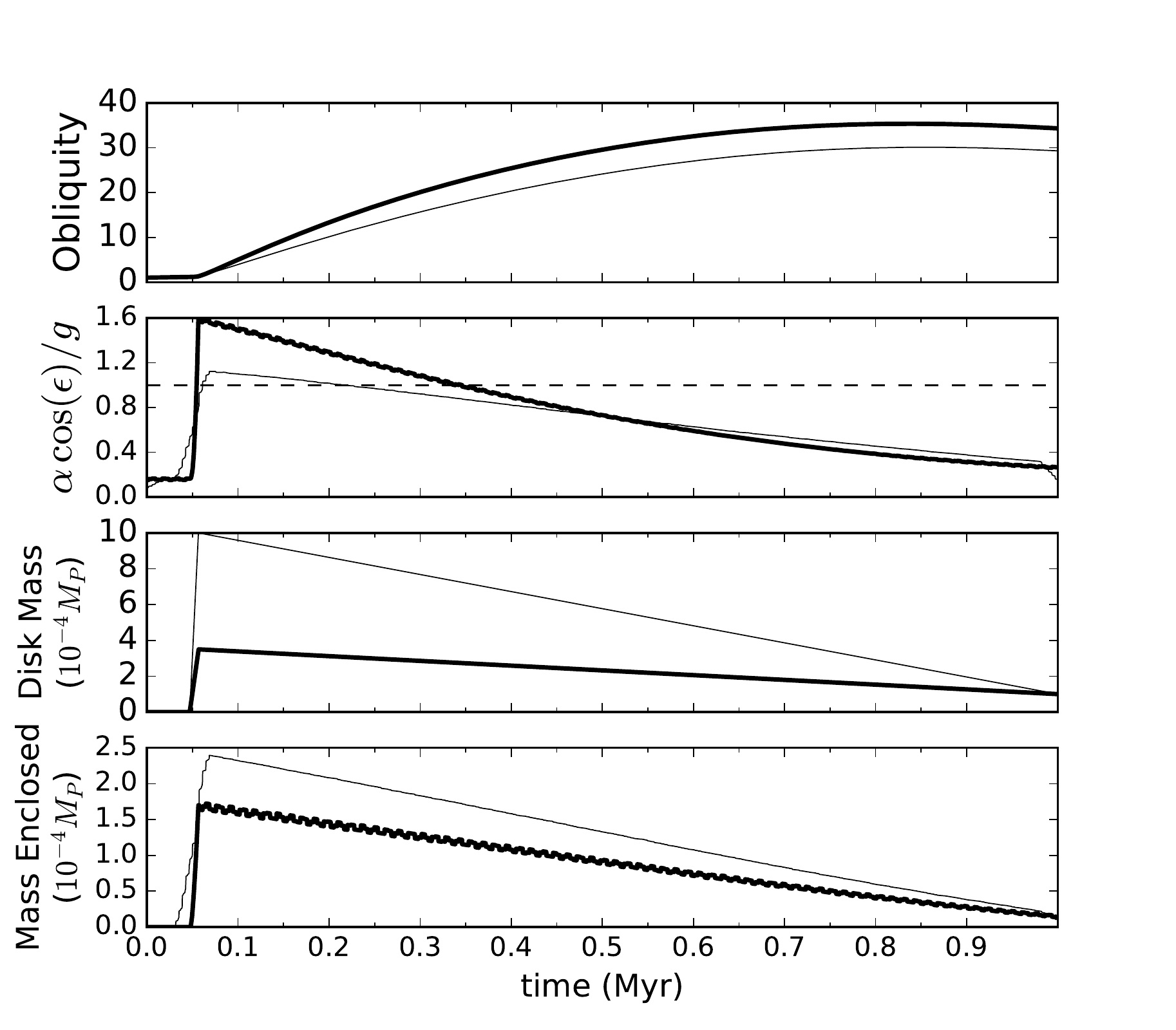}
	
	\caption{The evolution of Neptune's obliquity via a spin-orbit resonance if the planet harbored a massive disk. Here $M_{0}=0.9M_{N}$, $R_{0}=80R_{N}$, $a_{N}=28\,$au, $i_{N}=10$\textdegree, and Neptune's initial angular momentum is approximately the planet's current value. The thick bold lines have $R_{L}$ evolve according to Equation \ref{laplace2}, while the thin bold lines have $R_{L}$ depend only on the planet's quadrupole moment. The Laplace radius for the former case shrinks from $250-150\,R_{N}$, while in the latter case the Laplace radius increases from $180-195\,R_{N}$.}
	\label{fig:tilt-nep}
\end{figure}

Tilting Neptune is easier, since its obliquity needs only to be driven to 30\textdegree. If Neptune accreted its gas while located inside Uranus's current orbit in accordance with the Nice model \citep{2005Natur.435..466G,2005Natur.435..462M,2005Natur.435..459T} and grew similarly to Uranus as described previously, and we consider the two limiting scenarios for varying a planet's Laplace radius, then a disk with $M_{d}\approx7\times10^{-4}-4\times10^{-3}M_{N}$ can speed up its spin precession rate to generate a spin-orbit resonance and tilt Neptune assuming a primordial $i_{N}=3$\textdegree. Alternatively, if Neptune is located at 28 au with an inclination of 10\textdegree, then, as seen in Figure \ref{fig:tilt-nep}, the disk needs at least $3.5\times 10^{-4}M_{N}$ of gas to generate a spin-orbit resonance. The resonance drives Neptune's obliquity more weakly than Uranus's because libration rates are slower farther away from the Sun. In this figure, we set Neptune's initial spin angular momentum to be near its current value, and the disk's mass changes by only about 10\% if we reduce the planet's initial spin rate by a factor of 4. In the unphysical limiting case, where $R_{L}$ depends only on the planet's $J_{2}$, the disk needs to be twice as large to generate a resonance; regardless, a 30\textdegree~tilt can be attained in $\sim\!1$ Myr. If Neptune's inclination is instead 5\textdegree, then the accretion timescale needs to be 2 Myr to tilt the planet to $\sim30$\textdegree.

\section{Discussion}

Uranus and Neptune are not capable of entering a spin-orbit resonance today, as their spin axis precession rates are far too slow to match any of the planets' orbital precession frequencies. We have demonstrated that it is possible for both Uranus and Neptune to generate spin-orbit resonances if surrounded by a circumplanetary disk. Mass extending well beyond the classical Laplace radius can contribute to pole precession, meaning that the mass required to trigger a resonance is a modest 3-10 times the mass of their current satellite systems. Regardless of whether the disk remains in a steady state or is depleting, Uranus can be tilted up to 70\textdegree~if its orbit is inclined by more than 5\textdegree, and Neptune can be tilted all the way to 30\textdegree~with less inclined orbits. However, this strong resonance argument (Equation \ref{resarg}) is not capable of tilting planets beyond 90\textdegree~because the resonance will break as the planet's spin precession frequency nears zero (Equation \ref{period}). \cite{2018CeMDA.130...11Q} showed that a different resonant argument that includes mean motion terms and is not sensitive to orbital inclinations can push obliquities beyond 90\textdegree. This class of resonances requires additional planets potentially arranged in resonant chains. The forming giant planets may have started in or entered into such resonance chains, and in certain configurations, these mean-motion resonances can drive planets into a spin-orbit coupling \citep{2019NatAs...3..424M}. \cite{2003ApJ...597..566T} also argued that inclination growth can occur when planets are trapped into certain low order eccentricity-exciting mean-motion resonances, so an orbital evolution scenario that can simultaneously explain the configuration and tilts of the ice giants may exist. Ice giant formation models, however, do not require them to be placed into mean motion resonances as they acquire their gaseous atmospheres. There are a lot of potential scenarios, too many to pursue in this work. As for the cases discussed in this paper, we find that an additional collisional kick to Uranus's obliquity is inescapable.

\begin{figure}
	\centering
	\begin{tabular}[b]{@{}p{0.45\textwidth}@{}}
		\centering\includegraphics[width=0.4\textwidth]{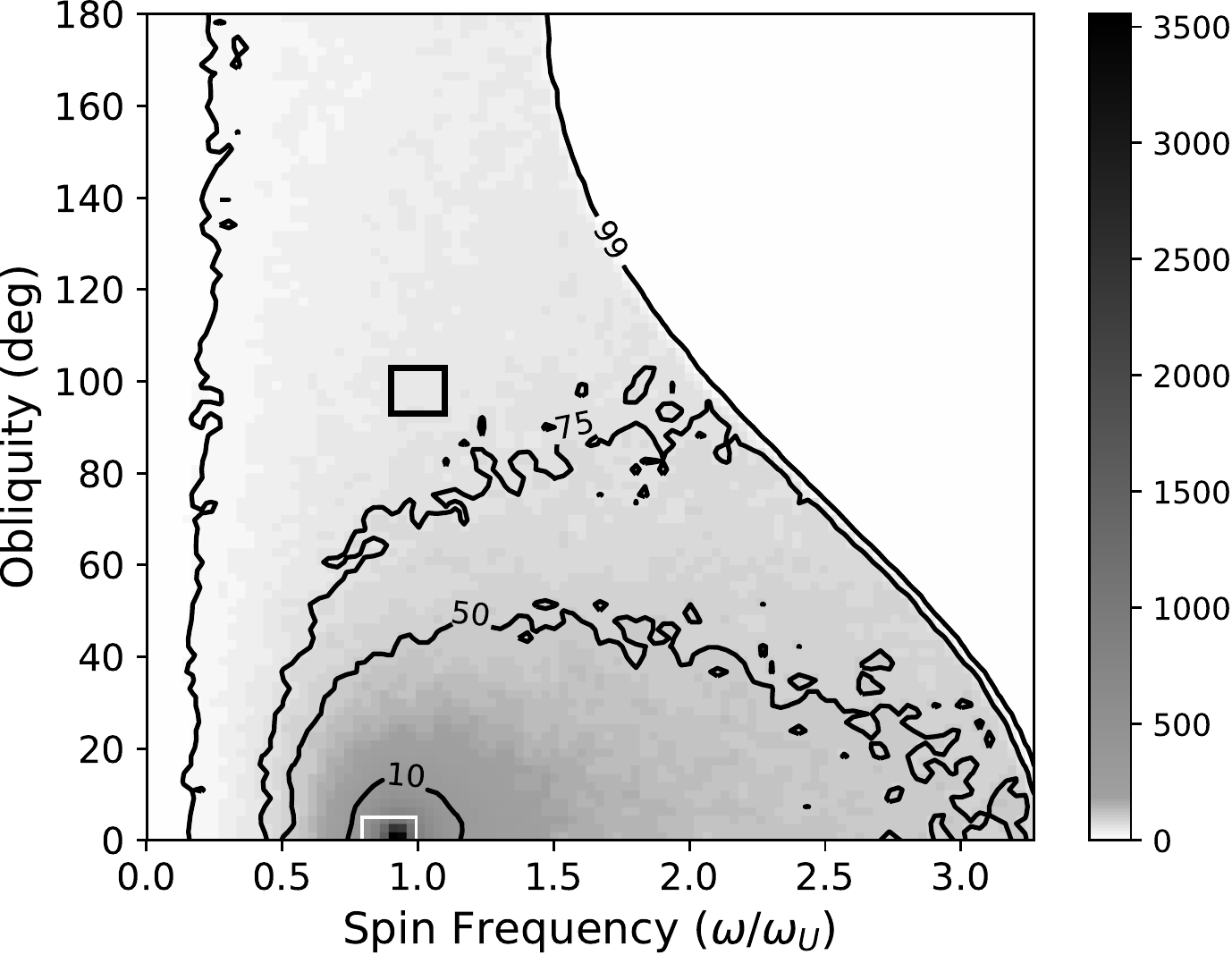} \\
		\centering (a)
	\end{tabular} \\
	\vspace{0.01cm}
	\begin{tabular}[b]{@{}p{0.45\textwidth}@{}}
		\centering\includegraphics[width=0.4\textwidth]{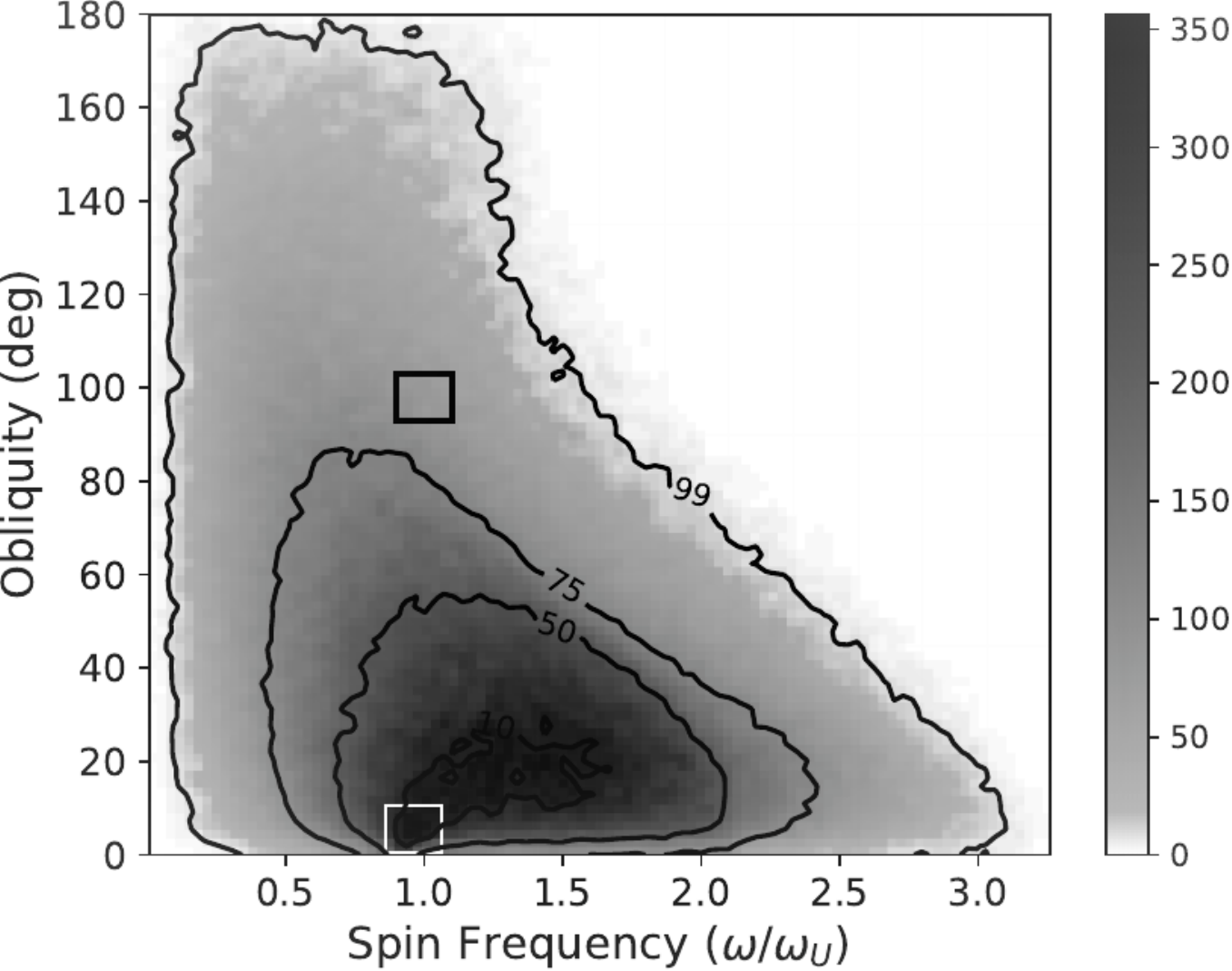} \\
		\centering (b)
	\end{tabular} \\
	\begin{tabular}[b]{@{}p{0.45\textwidth}@{}}
		\centering\includegraphics[width=0.4\textwidth]{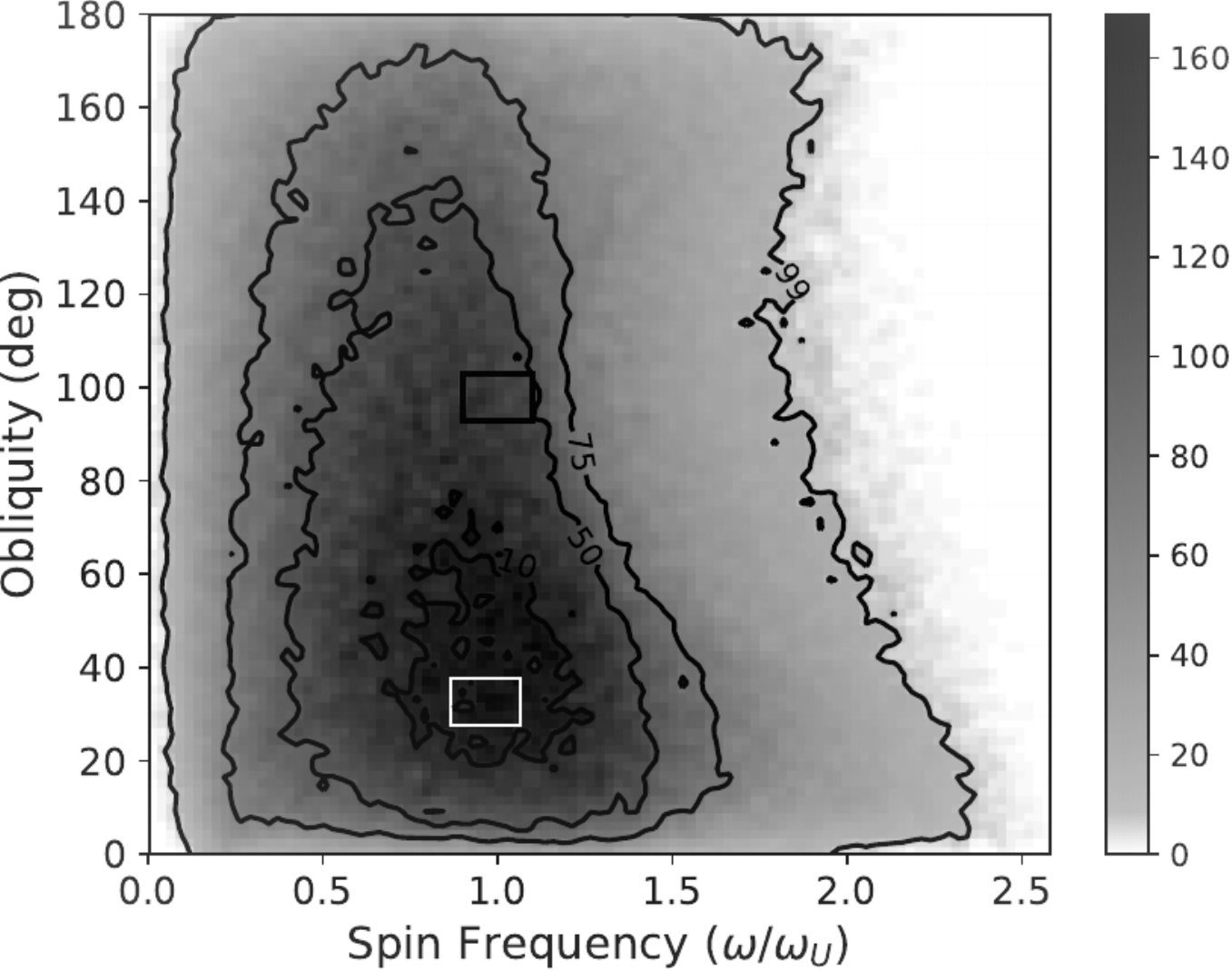} \\
		\centering (c)
	\end{tabular}
	
	\caption{ (a) Density plot of Uranus's obliquity and spin rate after a 1 $M_{\oplus}$ strike if its initial spin period is $T_{i}=16$ hr at $\epsilon_{i}=0$\textdegree~obliquity. Values within 10\% of Uranus's current obliquity and spin rate are contained inside the black box; the equivalent white box surrounds the peak of distribution. The color bar shows the number of instances for that value, and the contour lines contain the values within which a percentage of instances are found. The likelihood, $l$, of the planet's final spin state being within 10\% of its initial value is about 25 times greater than finding the planet within 10\% of Uranus's current spin state ($l_{U}$=0.0033). (b) Two 0.5 $M_{\oplus}$ strikes on a $T_{i}=16$ hr, $\epsilon_{i}=0$\textdegree~planet. The likelihood reduces to seven times more likely to find the planet near its initial value than within 10\% of Uranus's current state ($l_{U}$=0.0033). (c) Two 0.5 $M_{\oplus}$ strikes on a $T_{i}=68$ hr, $\epsilon_{i}=0$\textdegree~planet. Now, finding Uranus near its current value is only 1.3 times less likely ($l_{U}$=0.0098) than finding it near the maximum distribution.}
	\label{fig:impact-before}
\end{figure}

\begin{figure}
	\centering
	\begin{tabular}[b]{@{}p{0.45\textwidth}@{}}
		\centering\includegraphics[scale=0.5]{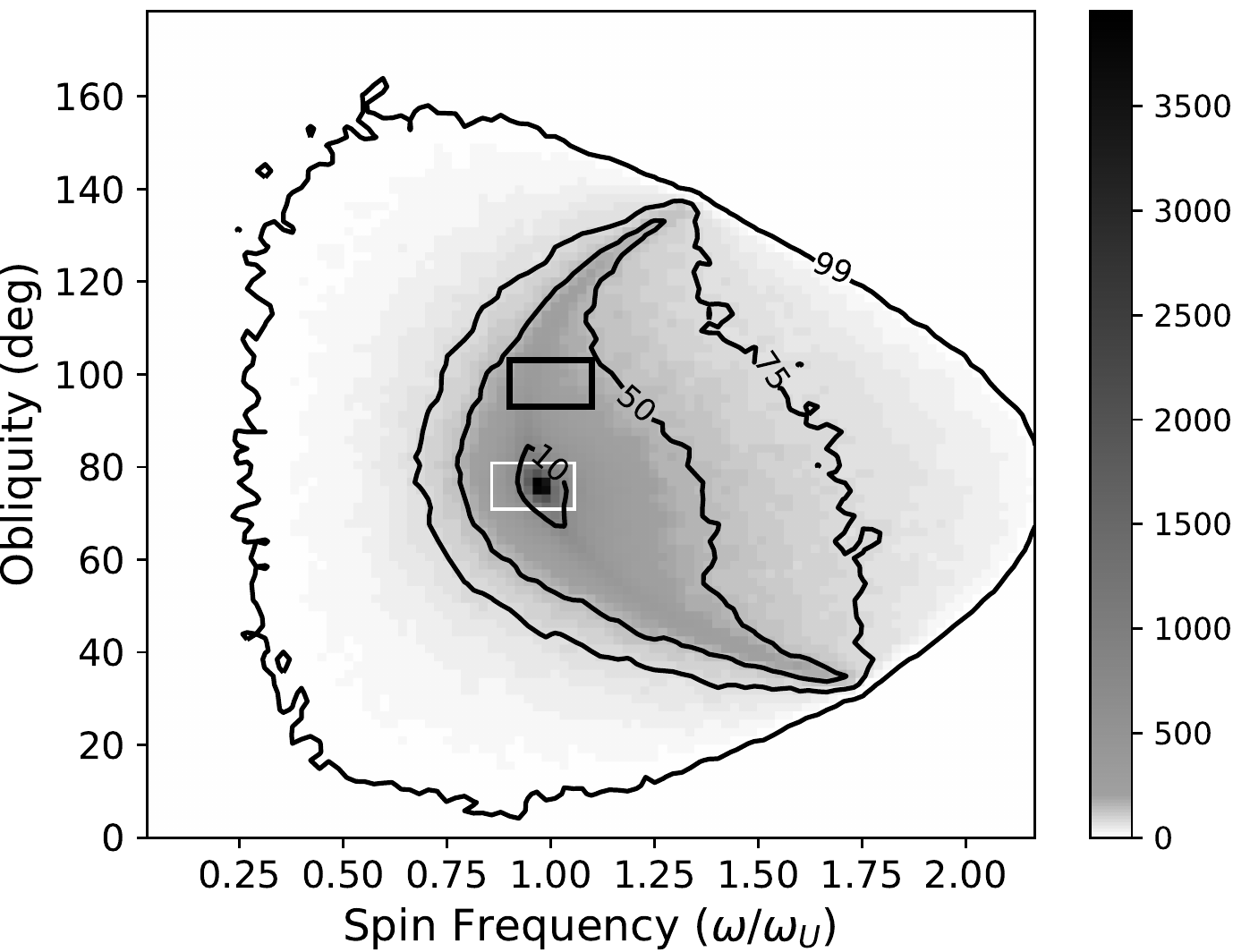} \\
		\centering (a)
	\end{tabular} \\
	\begin{tabular}[b]{@{}p{0.45\textwidth}@{}}
		\centering\includegraphics[scale=0.5]{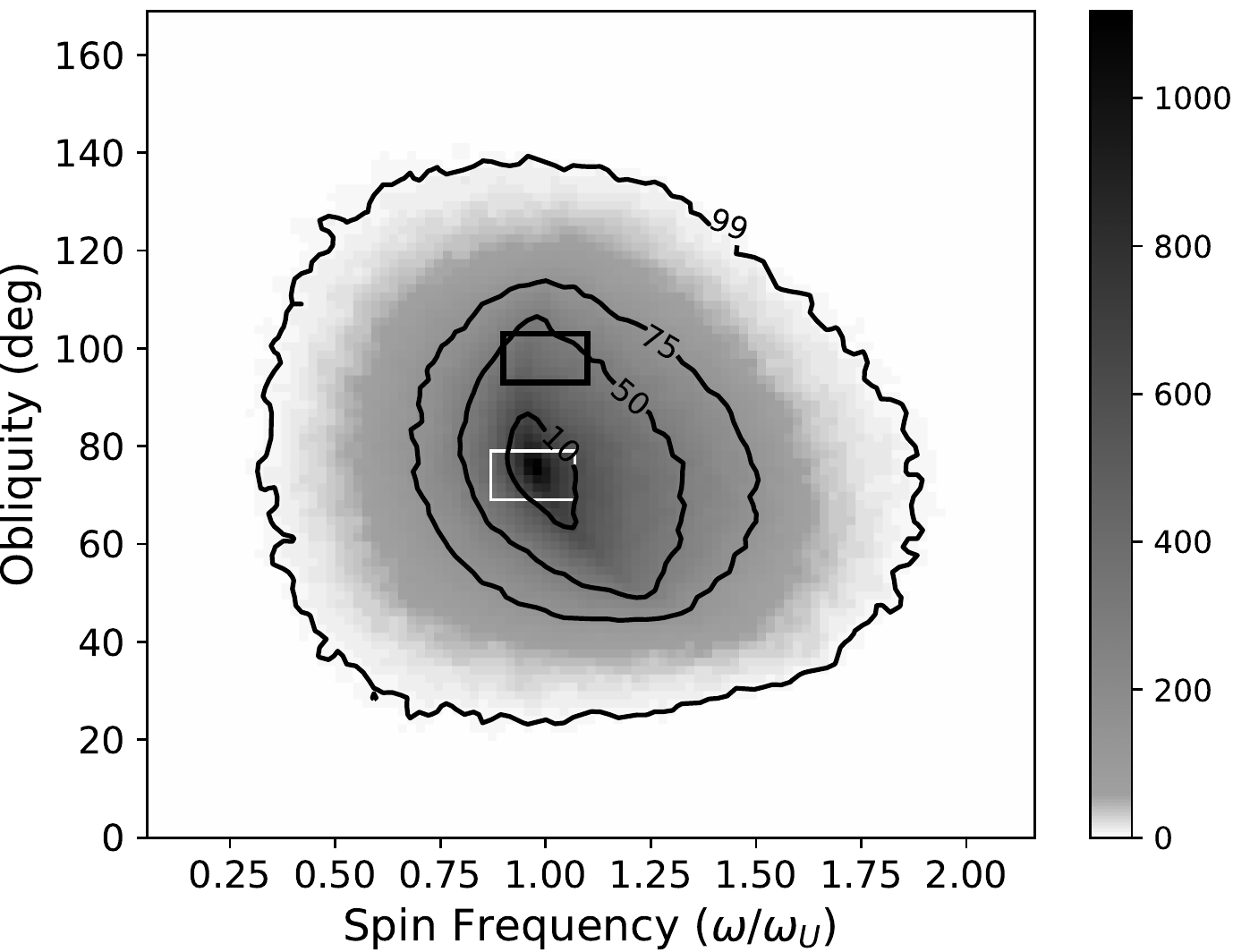} \\
		\centering (b)
	\end{tabular}
	
	\caption{Density plots of Uranus's obliquity and spin rate after a significant tilting. (a) Here $T_{i}=16$ hr and $\epsilon_{i}=75$\textdegree. Uranus is struck by one 0.5 $M_{\oplus}$ object. The likelihood, $l$, of the planet's final spin state being within 10\% of its initial value is 4.5 times greater than finding Uranus within 10\% of its current spin state ($l_{U}$=0.025). (b) Here, $T_{i}=16$ hr and $\epsilon_{i}=75$\textdegree. Uranus is struck by two 0.25 $M_{\oplus}$ objects. In this case, it is 2.1 times more likely to find the planet near the maximum value than finding Uranus within 10\% of its current spin state ($l_{U}$=0.038).}
	\label{fig:impact-after}
\end{figure}

If Uranus's and Neptune's spin periods are regulated entirely from gas accretion (Section \ref{sec:spin}), then these collisions cannot change their spin periods by more than about 10\%. Obliquities and spin periods, however, are each affected by collisions and are not independent variables. For instance, a normal strike to the equator will impart the most spin but will not tilt the planet. To quantify this, we developed a code that builds up a planet's spin by summing the angular momentum imparted by collisions striking random locations on the planet's surface for half a million realizations and calculates the planet's final spin state \citep{2019inprep}. Here we take into account gravitational focusing, as the planet's escape velocity is likely to be several times larger than the impactor's relative velocity on approach. For small relative velocities, where gravitational focusing is strong, then the impactor is focused significantly toward the planet's center:
\begin{equation}
b^{2}=R_{P}^{2}(1+(V_{esc}/V_{rel})^{2}).
\end{equation}
Here $b$ is the impact parameter, and the impactor approaches the planet on a hyperbolic orbit with speed $V_{rel}$ far from the planet. Since $V_{esc}^{2}=2GM_{P}/R_{P}$, $b^{2}\propto R_{P}$ for $V_{rel}\ll V_{esc}$. On the other hand, impactors striking the planet at very high velocities move on nearly straight lines and will instead yield a probability distribution proportional to the radius squared. We expect the impactors to approach the planet on initially eccentric elliptical orbits, and so we sample relative velocities between zero and 0.3 times Uranus's circular speed \citep{1994Sci...264..550H}.

Figure \ref{fig:impact-before} shows that a 1$M_{\oplus}$ collision will most likely not reproduce Uranus's current spin state if Uranus was initially spinning near its current rate. Since there is a higher concentration of radial impacts near the planet's center, the angular momentum imparted is small, and the distribution peaks strongly near the planet's initial state. Two strikes are an improvement, but we find better success if Uranus was initially spinning much slower than it is today. The odds of Uranus tilting to its current state for an initially slowly spinning planet is about an order of magnitude more likely than if it was initially spinning near its current rate. The mechanism responsible for removing a giant planet's angular momentum would then need to be more efficient for ice giants despite their more limited atmospheres, and as there is little justification for this, a pure giant collision scenario seems unlikely. 

This begs the question, though: how small can the planet's initial obliquity be such that a single impact can tilt the planet to 98\textdegree~with minimal variations to its spin period? Figure \ref{fig:impact-after} shows that Uranus's initial obliquity would need to be about 75\textdegree~to generate statistics as favorable as that for an initially slowly spinning planet. This also happens to be around the limit to which we can tilt Uranus with a spin-orbit resonance. The mass of the subsequent impactor would also need to be half as large ($0.5\,M_{\oplus}$), and the statistics even improve as the number of impactors increases to two $0.25\,M_{\oplus}$ objects (Figure \ref{fig:impact-after}b). Neptune's initial obliquity, on the other hand, would likely be zero, and its 30\textdegree~tilt could easily be a by-product of either a spin-orbit resonance, a single giant collision, or multiple giant collisions. 

Pebble accretion models predict an abundance of Mars-to-Earth-sized planets that have since disappeared \citep{2015Natur.524..322L,2015PNAS..11214180L}, so it is entirely possible that a few rogue planetary cores struck the ice giants. Our modeling shows that it is more probable, though, that the planets were struck by in total one of these objects rather than three or more. We believe that a hybrid model that includes both resonance and collisions is the most likely scenario, as it can eliminate the collision responsible for tilting Neptune, eliminates at least one of the impactors required to tilt Uranus \citep{2012Icar..219..737M}, and, most importantly, preserves the near equality of Uranus's and Neptune's spin rates. 

\section{Acknowledgement}
This work was supported by NASA Headquarters under the NASA Earth Science and Space Fellowship grant NNX16AP08H. The authors also thank Dr. Geoffrey Ryan for helpful discussions and an anonymous reviewer for useful advice, particularly on the Laplace radius.

\appendix
\section{Nodal Precession within a Protoplanetary Disk}\label{nodalprec}

Torques from neighboring planets cause a planet's orbit to precess. This precession rate is given as the sum of perturbations exterior and interior to the planet:
\begin{equation} \label{planet_orb_prec}
g_{+}\simeq -\frac{3}{4}\mu_{2}n_{1}\alpha^{3} \qquad\qquad \text{Exterior Perturber}\
\end{equation}
\begin{equation}\label{gmin}
g_{-}\simeq -\frac{3}{4}\mu_{1}n_{2}\alpha^{2} \qquad\qquad \text{Interior Perturber}\
\end{equation}
\citep{1999ssd..book.....M}. Here $\mu$ is the mass ratio of the perturber to the star, $n$ is the mean motion of the planet, and $\alpha=a_{1}/a_{2}$, where $a$ is the semimajor axis and the subscripts 1 and 2 refer to the inner and outer perturbers. We can transform these equations to instead describe perturbations from disks, as disks are made up of a series of concentric rings. For a surface density given by $\Sigma(r)=\Sigma_{0}(r/R_{o})^{-\beta}$, the mass of a protoplanetary disk can be described by \begin{equation}\label{density_prof}
M_{d} = \int_{R_{i}}^{R_{o}}\!\Sigma_{0}\left(\frac{r}{R_{o}}\right)^{-\beta} 2\pi rdr
\end{equation}
which can be integrated and solved for the constant reference surface density
\begin{equation}\label{surface_den}
\Sigma_{0} = \frac{\left(2-\beta\right)M_{d}}{2\pi \left(1-\eta^{2-\beta}\right)R_{o}^{2}}
\end{equation}
where $\eta=R_{i}/R_{o}$, $R_{i}$ is the inner radius of the disk, $R_{o}$ is its outer radius, and $\eta$ is always less than 1.

Setting $r_{p}$ as the planet--Sun distance that divides the interior and exterior disks, for an outer disk, we integrate Equation \ref{planet_orb_prec} radially over the disk, and we use Equation \ref{surface_den} to eliminate $\Sigma_{0}$. We set $R_{i}=a_{1}=r_{p}$ and integrate $r=a_{2}$ out to $R_{o}$ to find
\begin{equation}
g_{+} = -\frac{3}{4}\frac{2\pi\Sigma_{0}}{M_{\odot}}n_{1}r_{p}^{3} \int_{R_{i}}^{R_{o}}\!\left(\frac{r}{R_{o,+}}\right)^{-\beta}r^{-2}dr
\end{equation}
\begin{equation}
g_{+} = -\frac{3}{4}n\left(\frac{2-\beta_{+}}{-1-\beta_{+}}\right)\left(\frac{1-\eta_{+}^{-1-\beta_{+}}}{1-\eta_{+}^{2-\beta_{+}}}\right)\left(\frac{M_{d,+}}{M_{\odot}}\right)\left(\frac{r_{p}}{R_{o,+}}\right)^{3}.
\end{equation}
Similarly, for an interior disk, we use Equation \ref{gmin}, set $R_{o}=r_{p}=a_{2}$ and integrate $r=a_{1}$ from the inner boundary $R_{i}$ to find
\begin{equation}
g_{-} = -\frac{3}{4}\frac{2\pi\Sigma_{0}}{M_{\odot}}\frac{n_{2}}{r_{p}^{2}} \int_{R_{i}}^{R_{o}}\!\left(\frac{r}{R_{o,-}}\right)^{-\beta}r^{3}dr 
\end{equation}
\begin{equation}
g_{-} = -\frac{3}{4}n\left(\frac{2-\beta_{-}}{4-\beta_{-}}\right)\left(\frac{1-\eta_{-}^{4-\beta_{-}}}{1-\eta_{-}^{2-\beta_{-}}}\right)\left(\frac{M_{d,-}}{M_{\odot}}\right)\left(\frac{R_{o,-}}{r_{p}}\right)^{2}.
\end{equation}
Typically, we take $\beta_{-}=\beta_{+}$, but $M_{d,-}$ and $M_{d,+}$ can be quite different depending on the geometry. The expression for $g_{+}$ agrees with that obtained by \cite{2013ApJ...769...26C} using a different method, while to the best of our knowledge, that for $g_{-}$ is first given here.

\section{Laplace Radius with a Circumplanetary Disk}\label{laplacederiv}

Orbits located within a planet's Laplace radius precess about the planet's equator, while orbits located beyond the Laplace radius precess about the ecliptic plane. The transition between the two Laplace planes is gradual, and an approximation for this location is given as
\begin{equation}\label{full_Laplace}
R_{L} \approx \left(2J_{2,tot}\frac{M_{P}+M_{d}}{M_{\odot}}R_{P}^{2}r_{P}^{3}\right)^{1/5},
\end{equation}
where $J_{2,tot}=J_{2}+q$ is the total quadrupole of the planetary system, and $r_{P}$ is the planet's distance from the Sun. We can neglect $M_{d}$ since the mass of the circumplanetary disk or satellite system is usually much less than that of the planet, but the corresponding gravitational quadrupole moment is significant. The quadrupole moment of the Uranus's current satellite system is 4.7 times larger than the planet's $J_{2}$, and that value increases for an extended massive circumplanetary disk. 

A circumplanetary disk is composed of a series of nested massive rings, and those contained within the Laplace radius contribute to the disk's quadrupole moment. We can transform Equation \ref{quadrupole} by substituting the mass of the satellite with the mass of a ringlet, $dm=2\pi\Sigma(a)a\,da$, and replacing the summation with an integral. This gives
\begin{equation}
q=\int_{R_{P}}^{R_{L}}\!\frac{\pi\Sigma(a)}{M_{P}{R_{P}^{2}}}a^{3}\,da,
\end{equation}
where $a$ is the distance away from the central planet. In this derivation, we let the surface density profile of the disk fall as a power law,
\begin{equation}
\Sigma(a) = \Sigma_{0}\left(\frac{a}{R_{o}}\right)^{-\beta},
\end{equation}
where $\Sigma_{0}$ is the central surface density of the disk, $R_{o}$ is the outer radius of the disk, and $\beta>0$ is the power-law index. We typically compute the power-law index by assuming either a constant surface density or one that falls 3 orders of magnitude to the outer edge of the disk. The disk extends from the planet's surface to 0.3-0.5 Hill radii \citep{1998ApJ...508..707Q,2009MNRAS.397..657A,2012MNRAS.427.2597A,2009MNRAS.392..514M,2010AJ....140.1168W,2011MNRAS.413.1447M,2012ApJ...747...47T,2014ApJ...782...65S,2016ApJ...832..193Z}, but \cite{2018ApJ...868L..13S} focused specifically on ice giant formation models, and they focused their attention within 0.1 Hill radii. The disk's quadrupole moment can then be rewritten and solved assuming a Laplace radius much larger than the planet's radius
\begin{equation}\label{disk_q}
q=\frac{\pi\Sigma_{0}R_{o}^{\beta}}{M_{P}R_{P}^{2}}\int_{R_{P}}^{R_{L}}\!{a^{3-\beta}}da \approx \frac{\pi\Sigma_{0}R_{o}^{\beta}}{M_{P}R_{P}^{2}}\frac{R_{L}^{4-\beta}}{4-\beta}.
\end{equation}

If $q\gg J_{2}$, then substituting Equation \ref{disk_q} into Equation \ref{full_Laplace} gives
\begin{equation}
R_{L} \approx \left(\frac{2\pi\Sigma_{0}R_{o}^{\beta}r_{P}^{3}}{(4-\beta)M_{\odot}}\right)^{1/(1+\beta)}.
\end{equation}
    
\bibliography{bibliography}{}
\end{document}